\documentclass[useAMS,usenatbib,usegraphicx]{mn2e}
\usepackage{txfonts}
\usepackage{amscd}

\def\del#1{{}}

\newcommand{\ltsima}{$\; \buildrel < \over \sim \;$}
\newcommand{\lsim}{\lower.5ex\hbox{\ltsima}}
\newcommand{\gtsima}{$\; \buildrel > \over \sim \;$}
\newcommand{\gsim}{\lower.5ex\hbox{\gtsima}}

\newcommand{\bra}{\langle}
\newcommand{\ket}{\rangle}
\newcommand{\dang}{d_\mathrm{A}}

\newcommand{\dd}{\mathrm{d}}
\newcommand{\e}{\mathrm{e}}

\title[Morphological redshifts]
{Redshift estimation of clusters by wavelet decomposition of their Sunyaev-Zel'dovich morphology}

\author[B. M. Sch\"afer, C. Pfrommer and S. Zaroubi]
{B. M. Sch\"afer
\thanks{e-mail: spirou@mpa-garching.mpg.de (BMS); pfrommer@mpa-garching.mpg.de (CP); saleem@mpa-garching.mpg.de (SZ)},
C. Pfrommer\footnotemark[1] and S. Zaroubi\footnotemark[1] \\
Max-Planck-Institut f\"ur Astrophysik, Karl-Schwarzschild-Stra{\ss}e 1, Postfach 1317, 85741 Garching, Germany}

\begin{document}
\pagerange{\pageref{firstpage}--\pageref{lastpage}}
\pubyear{2003}
\maketitle
\label{firstpage}

\begin{abstract}
A method for estimating redshifts of galaxy clusters based solely on resolved Sunyaev-Zel'dovich (SZ) images is
proposed. Given a high resolution SZ cluster image (with FWHM of $\sim 1\arcmin$), the method indirectly measures its
structure related parameters (amplitude, size, etc.) by fitting a model function to the higher order wavelet
momenents of the cluster's SZ morphology. The applicability and accuracy of the wavelet method is assessed by applying
it to maps of a set of clusters extracted from hydrodynamical simulations of cosmic structure formation. The parameters,
derived by a fit to the spectrum of wavelet moments as a function of scale, are found to show a dependence on redshift
$z$ that is of the type $x(z) = x_1\exp(-z/x_2)+x_3$, where the monotony of this functional behaviour and the
non-degeneracy of those parameters allow inversion and estimation of the redshift $z$. The average attainable accuracy
in the $z$-estimation relative to $1+z$ is $\sim 4-5$\% out to $z\simeq1.2$, which is comparable to photometric
redshifts. For single-frequency SZ interferometers, where the ambient fluctuating CMB is the main noise source,
the accuracy of the method drops slightly to $\bra\Delta~z/(1+z)\ket\sim 6-7$\%.
\end{abstract}


\section{Introduction}\label{intro}
Inverse Compton scattering of cosmic microwave background (CMB) photons off thermal electrons within the hot
intra-cluster medium (ICM) of galaxy clusters produce fluctuations in the surface brightness of the CMB, an effect known
as the thermal Sunyaev-Zel'dovich (SZ) effect \citep[e.g.][]{1972SZorig,1980ARA&A..18..537S,1995ARA&A..33..541R}.
Imaging clusters of galaxies through their SZ signature has, until recently, been a very challenging undertaking. To
date, the development of detectors and new techniques have allowed high quality interferometric imaging of more than
fifty clusters of galaxies \citep{2002ARA&A..40..643C}, despite incomplete coverage of the Fourier plane. In the
foreseeable future, the availability of detectors in the microwave regime with angular resolutions surpassing $1\arcmin$
and sensitivities below $\mu\mathrm{K}$ \citep[e.g., the {\em South Pole Telescope}, described in detail in][]
{2002ARA&A..40..643C}, will probe the hot plasma in galaxy clusters out to large redshifts providing SZ based
wide field galaxy cluster catalogues and yielding a multitude of information about cluster formationand the cosmological
model \citep{1993birkinshaw}.

In particular, the abundance of clusters as a function of redshift has been shown to be a very sensitive probe of the
cosmological model \citep[][]{1998MNRAS.298.1145E,2000ApJ...534..565H}. The near independence of the
line-of-sight SZ amplitude on cluster redshift makes the SZ effect the main tool for detecting galaxy clusters at high
redshifts, $0.5\lsim z \lsim 2$ (the upper limit depends on cosmology quite sensitively). This range of redshifts is
especially important for probing the nature of the dark energy of the universe, since during this era it is expected to
evolve rapidly until it eventually dominates over the other cosmological fluids. In order to obtain precise constraints
on cosmological models it is essential to have accurate measurements of the redshift distribution of galaxy clusters
\citep[see][]{2001ApJ...553..545H}.

Normally, one determines the distance to the cluster by photometric or spectroscopic observations of the cluster member
galaxies. Unfortunately, this is a very challenging and time consuming task, in particular, when one considers the very
large number of mostly high redshift clusters expected to be observed with sensitive future SZ instruments -- The Planck
satellite alone is expected to detect about $10^4$ clusters \citep{2001A&A...370..754B}. In order to replace photometric
follow-ups we aim at inferring the distance to a cluster from SZ data alone for a future generation of experiments with
increased angular resolution of about $1\arcmin$.

Theoretically, the cold dark matter (CDM) hierarchical clustering paradigm predicts a universal profile for
dark matter halos that depends only on two parameters: core radius and density \citep{1995MNRAS.275..720N}.
In addition, the same theory provides a very simple recipe for the mass accretion history of a certain halo as a
function of its formation and observation redshift
\citep[][]{2002ApJ...568...52W,2002MNRAS.331...98V,2003MNRAS.339...12Z}. Using these relations together with simple
assumptions like hydrostatic equilibrium and isothermality, one can expect that in the framework of the spherical
collapse model the observable SZ flux and apparent size should provide measures of the cluster's mass and
distance.

Indeed, using scaling relations, \citet{2003MNRAS.341..599D} have demonstrated the viability of determining
reliable {\em morphological redshifts} and examined different SZ observables with respect to their distance sensitivity.
Among those observables, they showed that the cluster apparent size and central amplitude are promising distance
indicators, once their degeneracy is broken.

The main goal of this work is to derive redshifts of clusters based solely on their resolved SZ images by modeling the
evolution of their structural parameters with redshift from the data set itself. This phenomenological approach does not
depend on a priori assumptions about scaling relations that are valid only for spherically symmetric and relaxed
systems.

Specifically, the structural morphology of the cluster's pressure profile in an SZ observation is characterised by
wavelet analysis.\footnote{There are also various ways of characterising the cluster's density profile in an SZ
observation that are more or less susceptible to noise, for instance the fitting of a $\beta$-profile
\citep{1978A&A....70..677C} to the electron density.}  We are able to show that there is a simple relation between
the distribution of moments over various scales in wavelet space and the cluster properties which can be described with
simple phenomenological functions. Furthermore, the parameters of these functions are shown to follow a well defined and
simple redshift dependence. Wavelet analysis has been chosen because it maintains the scale and positional information
of cluster morphology, hence, it makes isolation and suppression of various unwanted contributions to the
observed signal possible while it reliabely upholds the underlying behavior. We note however, that Fourier space
analysis could in principle yield very similar results.

Hydrodynamically simulated clusters are used to demonstrate the method and to set limits on the redshift uncertainty
expected in this approach. The simulated clusters used in the analysis are close to virialisation, e.g. merging systems
are excluded. Under this restriction, both the relation between the observed quantity and the cluster physical
parameters as well as the structural parameters are well defined. In addition, simulated clusters ignore radiative
and feedback processes, the effect of which is discussed later in the paper.

In the observational application, the evolution of the structural parameters following from wavelet decomposition could
be calibrated from a (relatively small) learning set of high quality SZ clusters with known (photometric/spectroscopic)
redshifts.

Our method relies crucially on the availability of resolved SZ cluster images. Therefore, throughout the paper we assume
an instrumental resolution of $1\arcmin$, where massive clusters should be resolved even at the largest
redshifts considered here.  Indeed, future instruments such as the South Pole
Telescope\footnote{{\tt http://astro.uchicago.edu/spt/}} \citep{2002ARA&A..40..643C} or the
Atacama Cosmology Telescope\footnote{{\tt http://www.hep.upenn.edu/{\textasciitilde}angelica/act/act.html}} are
designed to yield observations of up to $10^4$ galaxy clusters with masses $\gsim 10^{14} M_\odot$
($1\mbox{ }\mu\mbox{K}$ sensitivity) and $\approx 1\arcmin$ resolution.

This article is organised as follows: After basic definitions concerning the SZ effect in Sect.~\ref{szdef} and
wavelets in Sect.~\ref{wavelet}, the simulations are outlined in Sect.~\ref{sim}. The capability of wavelets with
respect to distance estimation is examined in Sect.~\ref{ana}. Possible systematics are addressed in
Sect.~\ref{systematics}. A summary of the techniques in Sect.~\ref{nutshell} and of the results in Sect.~\ref{sum}
concludes the article.

\section{Sunyaev-Zel'dovich definitions}\label{szdef}
The SZ effect has been described in detail by many authors \citep[for a comprehensive review see][]
{1993birkinshaw}; here we briefly review its main aspects. The SZ effect arises because CMB photons experience
Compton-scattering off electrons of the dilute intra-cluster plasma. The CMB spectrum is modulated as photons are
redistributed from the low-frequency part of the spectrum below $218\mbox{ GHz}$ to higher frequencies. The change in
thermodynamic CMB temperature due to the thermal SZ effect is
\begin{equation}
\frac{\Delta T}{T}(\bmath{\phi})=
y(\bmath{\phi})\,\left(x\frac{e^x+1}{e^x-1}-4\right)\simeq-2y(\bmath{\phi})\quad\mbox{for $x\ll1$,}
\label{sz_temp_decr}
\end{equation}
where $x=h\nu/k T_\mathrm{CMB}$  is the dimensionless frequency. In the Rayleigh-Jeans limit ($x\ll 1$), the change in
temperature is asymptotically equal to $-2y(\bmath{\phi})$. The SZ amplitude at location $\bmath{\phi}$, which is known
as the Comptonisation parameter $y(\bmath{\phi})$, is defined as the line-of-sight integral of the temperature-weighted
thermal electron density:
\begin{equation}
y(\bmath{\phi}) = \frac{\sigma_\mathrm{T} k}{m_\e c^2}\int\dd l\:n_\e(\bmath{\phi},l)T_\e(\bmath{\phi},l)\mbox{.}
\end{equation}
where $m_\e$, $c$ and $k$ denote electron mass, speed of light and Boltzmann's constant, respectively.
$T_\e(\bmath{\phi},l)$ and $n_\e(\bmath{\phi},l)$ are electron temperature and electron number density at
position $\bmath{\phi}$ and distance $l$.

\section{Wavelets}\label{wavelet}\label{def_wavelet}

\subsection{Wavelet definitions}\label{definitions}
During the last decade, wavelet analysis has become a popular tool in various data analysis and image processing
applications. The main appeal of wavelet functional bases stems from their simultaneous localisation of a signal in both
the wavenumber and position domain, where they make orthogonal and complete projections on modes belonging to both
spaces possible. In particular, the discrete wavelet families, by virtue of their constituting a complete basis,
provide a unique and fast decomposition of the images into wavelet expansion coefficients. Statistics in terms of
the $q^\mathrm{th}$ moments of the distribution of wavelet coefficients as a function of scale compresses the
signal contained in an image into a small number of parameters and yields information surpassing that derived in
traditional Fourier analysis.

Following \citet{1993ASAJ...93.1671D} and \citet{1993PhRvE..47..875M}, the wavelet transform of a 2-dimensional image is
defined as a convolution of the function $y(\bmath{x})$ to be analysed with the wavelet
$\psi_\sigma\left(\left|\bmath{x}-\bmath{\mu}\right|\right)$:
\begin{equation}
\chi(\bmath{\mu},\sigma)
=\int\dd^2x\:y(\bmath{x})\cdot\:\psi_\sigma\left(\left|\bmath{x}-\bmath{\mu}\right|\right)\mbox{.}
\label{wavelet_expand}
\end{equation}
High values for $\chi(\bmath{\mu},\sigma)$ are obtained in case of a match between the features of $y(\bmath{x})$ and
the wavelet $\psi_\sigma(\bmath{x})$ at position $\bmath{\mu}$ and scale $\sigma$. From the wavelet expansion
coefficients $\chi(\bmath{\mu},\sigma)$ on scale $\sigma$ at location $\bmath{\mu}$ one obtains the wavelet moments
$X_q(\sigma)$ by integration over all displacements $\bmath{\mu}$:
\begin{equation}
X_q(\sigma)=\int\dd^2\mu\,\left|\chi\left(\bmath{\mu},\sigma\right)\right|^q\mbox{.}
\label{wavelet_spec}
\end{equation}
The exponent $q\in\mathbb{N}$ defines the order of the wavelet moment $X_q(\sigma)$. Values for $q$ equal or larger than
2 allow noise suppression. The logarithm $\ln X(\sigma,q)$ of the wavelet moment as a function of logarithmic scale
$\ln\sigma$ constitutes the wavelet spectrum. The $X_q(\sigma)$-statistic is the main tool used in this study for
characterising the morphology of SZ clusters.

\subsection{Application of wavelets to a cluster profile}
\subsubsection{Analytic wavelet transform of a cluster $y$-profile}\label{application}
In order to illustrate our idea of determining cluster sizes via wavelet decomposition, the wavelet transform of a King
profile, which is known to describe the SZ morphology of clusters to first order, is performed. As an analysing wavelet,
the Mexican-hat wavelet was chosen for simplicity.

It is favourable to compute the convolution in the definition of $\chi(\bmath{\mu},\sigma)$ in the Fourier domain. By
virtue of eqn.~(\ref{conv_theorem}),
\begin{eqnarray}\chi(\bmath{\mu},\sigma) & = &
\int\dd^2x\:y(\bmath{x})\:\psi_\sigma\left(\bmath{x}-\bmath{\mu}\right) \\ & =
&(2\pi)^2\int\dd^2k\:Y(\bmath{k})\:\Psi_\sigma(-\bmath{k})\exp(\mathrm{i}\bmath{k}\bmath{\mu})\mbox{,}
\label{conv_theorem}
\end{eqnarray}
the convolution reduces to a mere multiplication of the Fourier transforms $Y(\bmath{k})$ and $\Psi_\sigma(\bmath{k})$
of the image $y(\bmath{x})$ and the wavelet $\psi_\sigma(\bmath{x})$, respectively. Restricting the order of the
wavelet moment to $q = 2$ and inserting the convolution theorem~(\ref{conv_theorem}) into the
definition~(\ref{wavelet_spec}) yields:
\begin{eqnarray}X_2(\sigma) 	& = &
(2\pi)^4\int\dd^2\mu\:\left|\int\dd^2k\:Y(\bmath{k})\Psi_\sigma(-\bmath{k})\exp(\mathrm{i}\bmath{k}\bmath{\mu})\right|^2
\\& =
&(2\pi)^6\int\dd^2k\left|Y(\bmath{k})\right|^2\left|\Psi_\sigma(\bmath{k})\right|^2\mbox{,}
\label{centre_form}
\end{eqnarray}
where the replacement $\left|\Psi_\sigma(-\bmath{k})\right|^2 = \left|\Psi_\sigma(\bmath{k})\right|^2$ holds for
real wavelets.

The Mexican-hat wavelet is defined as the negative Laplacian of a Gaussian:
\begin{equation}
\psi_\mathrm{MH}(\bmath{x})=\psi_\mathrm{MH}(x)=-\nabla_{\bmath{x}}^2\left[\frac{1}{2\pi\sigma
^2}\exp\left(-\frac{\bmath{x}^2}{2\sigma^2}\right)\right]\mbox{,}\label{psi}
\end{equation}
whereof the Fourier transform $\Psi_\mathrm{MH}(\bmath{k})$ is derived by twofold partial integration:
\begin{eqnarray}
\Psi_\mathrm{MH}(k) & = & \int\frac{\dd^2x}{(2\pi)^2}\:\psi_\mathrm{MH}(\bmath{x})\exp(-\mathrm{i}\bmath{k}\bmath{x}) \\
& = & \frac{1}{(2\pi)^2\sigma^6}\int r\dd r (2\sigma^2-
r^2)\exp\left(-\frac{r^2}{2\sigma^2}\right)J_0(kr)\\&=&\frac{k^2}{(2\pi)^2}\cdot\exp\left(-\frac{k^2\sigma^2}{2}\right)
\mbox{,}
\label{fftpsi}
\end{eqnarray}
where the azimuthal symmetry and the definition of the zeroth order Bessel function of the first kind, $2\pi
J_0(kr)=\int_0^{2\pi}\dd\phi\exp\left(\mathrm{i}kr\cos\phi\right)$ was used in the first step. Thus, the
Fourier transform of the wavelet, $\Psi_\sigma(k)$, is given by the Hankel transform of the Laplacian of a Gaussian.

For the determination of $Y(\bmath{k})$, we assume that the projected thermal electron density can be described by a
spherically symmetric King profile, i.e. a $\beta$-model \citep{1978A&A....70..677C} with $\beta = 1$, core radius $r_c$
and central value of the Comptonisation parameter $y_0$:
\begin{equation}y(\bmath{x}) = y(r)
=y_0\left[1+\left(\frac{r}{r_c}\right)^2\right]^{-1}\mbox{.}\label{beta_profile}
\end{equation}
Then, the Fourier transform is easily computed:
\begin{eqnarray}Y(k) 	& = & \int\frac{\dd^2
x}{(2\pi)^2}\:y(\bmath{x})\exp(-\mathrm{i}\bmath{k}\bmath{x}) \\&= & \frac{y_0 r_c^2}{2\pi}\int\dd
r\:\frac{r}{r_c^2+r^2}J_0(kr) =\frac{y_0 r_c^2}{2\pi}\cdot K_0(kr_c)\mbox{,}\label{ffty}
\end{eqnarray}
where in eqn.~(\ref{ffty}) the definition of the zeroth order modified Bessel function of the second kind $K_0(k r_c)$
was inserted.

Substituting eqns.~(\ref{fftpsi}) and (\ref{ffty}) into eqn.~(\ref{centre_form}) and exploiting the azimuthal symmetry
of the functions $y(\bmath{x})$ and $\psi(\bmath{x})$ yields an analytic integral for $X_2(\sigma)$:
\begin{equation}
X_2(\sigma) = 2\pi y_0^2r_c^4\int_0^{\infty}\dd k\:k^5\exp\left(-\sigma^2 k^2\right)
K_0^2(k r_c)\mbox{.}
\label{ana_wave_trafo}
\end{equation}
After evaluation of the integral in eqn.~(\ref{ana_wave_trafo}), the wavelet transform of the $\beta$-profile reads
as follows:
\begin{eqnarray}
X_2(\sigma) & =
&\frac{\pi^{3/2}y_0^2}{2r_c^2}\;\alpha^6\cdot{\displaystyle\mathcal{G}^{3,1}_{2, 3}}\left({\textstyle\alpha^2}
\big|{\textstyle { -2\;\;\frac{1}{2} \atop 0\, 0\, 0}}\right)\mbox{,}
\label{analytic_spec}
\end{eqnarray}
where $\alpha=r_c/\sigma$ has been substituted. The function ${\displaystyle\mathcal{G}}$ is Meijer's G-function, the
exact definition of which is given by \citet{1994tisp.book.....G}. It is an interesting consistency to note that apart
from the normalisation, the functional shape of eqn.~(\ref{analytic_spec}) only depends on $\alpha$, i.e. on the core
radius $r_c$ expressed in units of the wavelet scale $\sigma$.

\subsubsection{Asymptotics of the analytical wavelet transform}\label{asymptotics}
The asymptotic behavior of $X_2(\sigma)$ at the limit of $\sigma\ll r_c$ can be explored by substituting the
expressions given in eqns.~(\ref{beta_profile}) and (\ref{psi}) into eqn.~(\ref{conv_theorem}), and exchanging, by
partial integration, the function on which the Laplacian operates. In the limit of interest the Gaussian can be
replaced by a Dirac-$\delta$ distribution. Substituting all of this into eqn.~(\ref{wavelet_spec}) yields
that $\lim_{\sigma\to0}X_2(\sigma)$ is proportional to $y_0^2$ and independent of $\sigma$, i.e. the normalisation of
the wavelet spectrum measures the square of the central Comptonisation parameter $y_0$:
\begin{equation}
X_2(\sigma)=\frac{32\pi}{15}\cdot\frac{y_0^2}{r_c^2}\qquad\mbox{for}\qquad\sigma\ll r_c\mbox{.}
\end{equation}
In the opposite limit, i.e. $r_c \ll \sigma$, one can use the fact that the King-profile is highly peaked at the
center and that it is convolved with a Mexican-hat wavelet guaranteeing the convergence of the integral in
eqn.~(\ref{ana_wave_trafo}) at $\infty$. In the limit of $r_c\rightarrow 0$ this integral is dominated by the value at
$k=0$. Therefore, one can approximate the King-profile with a Dirac-$\delta$ distribution and show that
asymptotically the $\lim_{r_c \to 0}X_2(\sigma)$ is proportional to $\sigma^{-6}$:
\begin{equation}
X_2(\sigma)\propto\frac{y_0^2r_c^4}{\sigma^{6}}\qquad\mbox{for}\qquad\sigma\gg r_c\mbox{.}
\end{equation}

The sensitivity of the wavelet spectrum $X_2(\sigma)$ on cluster size $r_c$ is illustrated by
Fig.~\ref{conv_theorem}. The wavelet spectrum is constant for $\sigma\ll r_c$, has an $r_c$-dependent break and
drops off asymptotically $\propto\sigma^{-6}$ for $\sigma\gg r_c$. Naturally, the scale $\sigma$ at which the
transition from one asymptotic regime to the other occurs, is determined by the value of $r_c$, i.e. the cluster size.

\begin{figure}
\resizebox{\hsize}{!}{\includegraphics{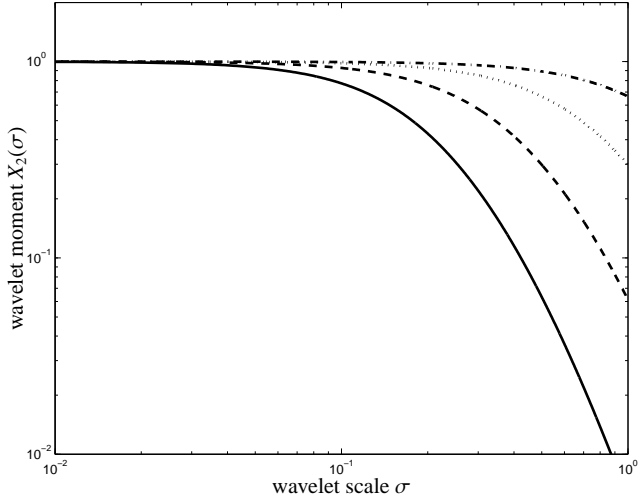}}
\caption{Sensitivity of the wavelet spectrum on the cluster size: The second order wavelet moments $X_2(\sigma)$ are
shown as a function of $\sigma$ for various core sizes $r_c = 0.5$ (solid), $r_c = 1$ (dashed), $r_c = 2$ (dotted) and
$r_c =4$ (dash-dotted). The curves have been normalised to their asymptotic values for $\sigma\rightarrow 0$.}
\label{fig_analytic_spec}
\end{figure}

Motivated by this example, the wavelet moments $X_q(\sigma)$ obtained from real data (Sect.~\ref{ana_wavelet_spec})
will be fitted with a power law with an exponential cutoff, where the cutoff indicates the cluster size and the
amplitude is proportional to some power of the central Comptonisation parameter $y_0$.

\subsubsection{Finite instrumental resolution}\label{finite_resolution}
The influence of finite instrumental resolution can easily be incorporated by an additional factor $\left|B(k)\right|^2$
in eqn.~(\ref{centre_form}):
\begin{equation}
X_2(\sigma) =(2\pi)^6\int\dd^2 k\left|Y(k)\right|^2 \left|\Psi_\sigma(k)\right|^2
\left|B(k)\right|^2\mbox{,}
\label{variance_with_beam}
\end{equation}
where $B(k)$ is the Fourier transform of the (azimuthally symmetric) beam profile $b(x)$, which is for simplicity
assumed to be of Gaussian shape with FWHM~$=\sqrt{8\ln(2)}\cdot\sigma_b$:
\begin{eqnarray}
B(k) & = & \int\frac{\dd^2
x}{(2\pi)^2}\:b(x)\exp(-\mathrm{i}\bmath{k}\bmath{x}) \mbox{ with}\\b(\bmath{x}) & =
&\frac{1}{2\pi\sigma_b^2}\exp\left(-\frac{\bmath{x}^2}{2\sigma_b^2}\right)\mbox{.}
\end{eqnarray}
This effectively replaces $\sigma$ in eqn.~(\ref{analytic_spec}) by the harmonic mean $\sqrt{\sigma^2+\sigma_b^2}$,
which limits the range of accessible wavelet scales to $\sigma > \sigma_b$.

\subsection{Analogy to power spectra in Fourier analysis}\label{analogy}
By interpreting the wavelet spectrum in eqn.~(\ref{centre_form}) as the variance of the fluctuations on the scale
$\sigma$, one may draw an analogy to Fourier decomposition:
\begin{equation}
\mathrm{var}\left[y(x)\right] =
X_2(\sigma)=(2\pi)^4\int\dd^2 k\:P(k)\left|\Psi_\sigma(k)\right|^2\mbox{,}
\end{equation}
where $P(k) = (2\pi)^2\bra\left|Y(k)\right|^2\ket$ is the Fourier power spectrum. The wavelet $\psi(x)$ now adopts the
role of a filter function on scale $\sigma$. This filter function reads in real space in the case of the Mexican hat
wavelet:
\begin{equation}
\psi_\mathrm{MH}(\bmath{x})=\frac{2\sigma^2-\bmath{x}^2}{2\pi\sigma^6}\cdot\exp\left(-\frac{\bmath{x}^2}{2\sigma^2}
\right)\mbox{.}
\end{equation}
Therefore, our method is equivalent to considering power spectral analysis of filtered fields and higher order
Fourier space moments.

\section{Simulations}\label{sim}
The accuracy in the determination of redshift $z$ was assessed by examining the performance on numerical
simulations: First, simulations of cosmological structure formation including gas physics have been carried out in order
to model the evolution of clusters (Sect.~\ref{sim_hydro}). Subsequently, maps of the Compton-$y$ parameter have been
produced by using an interpolation kernel with an adaptive smoothing length for projecting the Compton-$y$ parameter
along the line-of-sight (Sect.~\ref{sim_project}). By applying selection criteria favouring virialised systems a cluster
sample was compiled (Sect.~\ref{selection_criteria}). Finally, aiming at realistic single frequency SZ observations, we
simulated the ambient CMB fluctuations that act as the primary source of noise (Sect.~\ref{cmb_fast_descr}) and combined
the resulting realisations of the CMB with the cluster maps (Sect.~\ref{sim_obs_descr}).

The assumed cosmological model is the standard \mbox{$\Lambda$CDM} cosmology, which has recently  been supported by the
WMAP satellite \citep[][]{2003astro.ph..2207B,2003astro.ph..2209S}. Parameter values have been chosen as
$\Omega_\mathrm{M} = 0.3$, $\Omega_\Lambda =0.7$, $H_0 = 100\,h\,\mbox{km}\mbox{s}^{-1}\mbox{ Mpc}^{-1}$ with $h = 0.7$,
$\Omega_\mathrm{B} = 0.04$, $n_\mathrm{s} =1$ and $\sigma_8=0.9$.

\subsection{SPH cluster simulations}\label{sim_hydro}
A simulation of cosmological structure formation kindly provided by V. Springel and L. Hernquist
\citep[][]{2002MNRAS.333..649S,2002ApJ...579...16W} constitutes the basis of our analysis. In a cubical box of
comoving side length $100\mbox{ Mpc}/h$ with periodic boundary conditions a smoothed particle hydrodynamic (SPH)
simulation comprising $216^3$ dark matter particles as well as $216^3$ gas particles was run and snapshots were saved at
23 redshifts ranging from $z =0.102$ out to $z = 1.114$. The comoving spacing along the line-of-sight of two subsequent
boxes is $100\mbox{ Mpc}/h$. Purely adiabatic gas physics and shock heating were included, but radiative cooling and
star formation were ignored, which however does not result in significant differences in SZ morphology, as shown by
\citet{2002ApJ...579...16W} but impacts on the scaling relations as demonstrated by \citet{2001ApJ...561L..15D}.

Overdensities are identified using a friends-of-friends algorithm with the linking length $b= 0.164$, which yields all
member particles of a cluster in conjunction with a spherical overdensity code, from which virial quantities are
estimated. We computed the mass $M_\mathrm{vir}$ inside a sphere of radius $r_\mathrm{vir}$, interior to which the
average density was 200 times the critical density $\rho_\mathrm{crit}=3H_0^2/(8\pi G)$. The angle subtended by twice
the virial radius is denoted as $\theta_\mathrm{vir}$. We imposed a lower mass threshold of $M_\mathrm{vir}\geq 5\cdot
10^{13}M_{\sun}/h$.

\subsection{SZ map preparation}\label{sim_project}
\begin{figure*}
\begin{tabular}{ccc}
\resizebox{5.5cm}{!}{\includegraphics{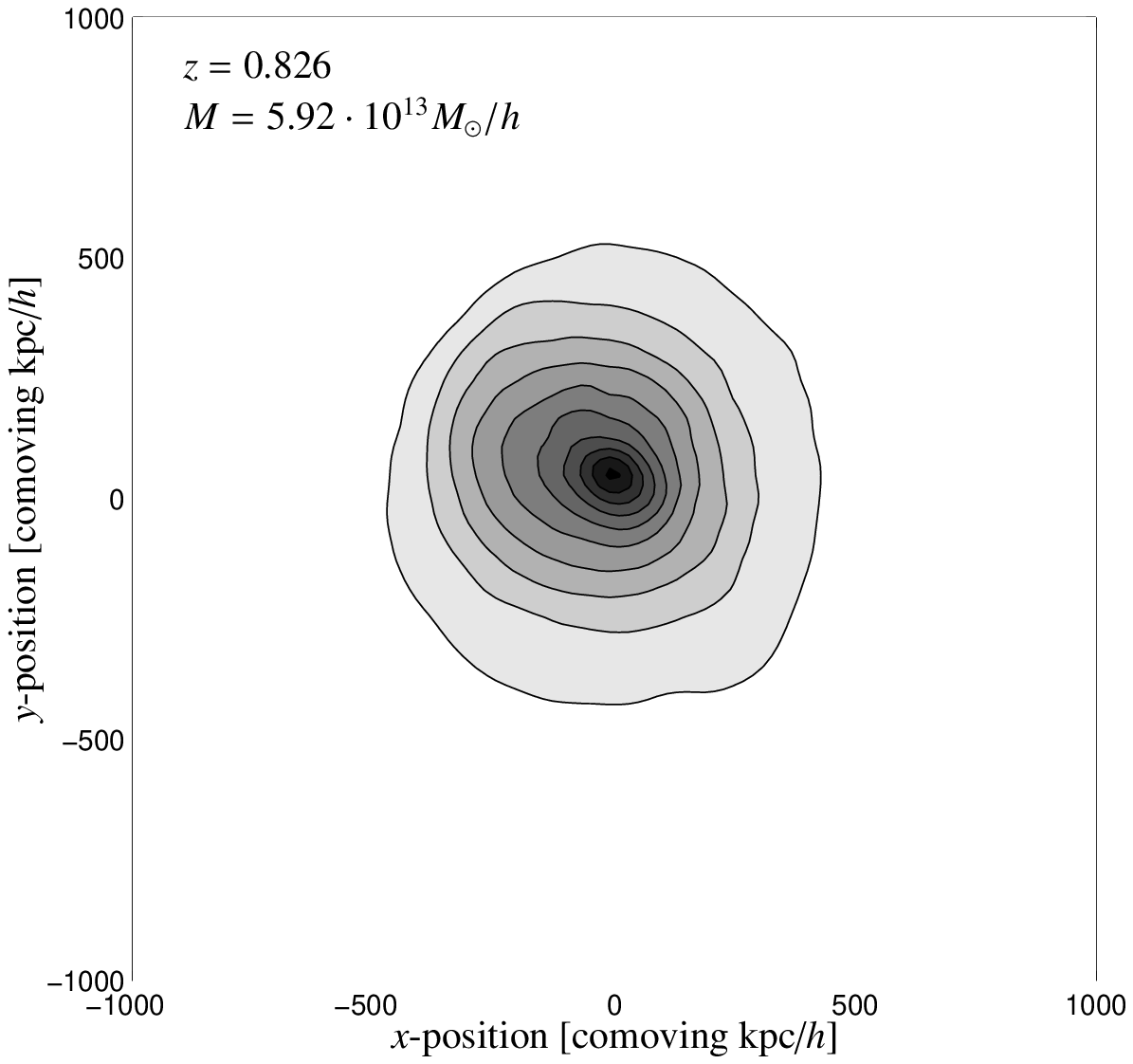}} &
\resizebox{5.5cm}{!}{\includegraphics{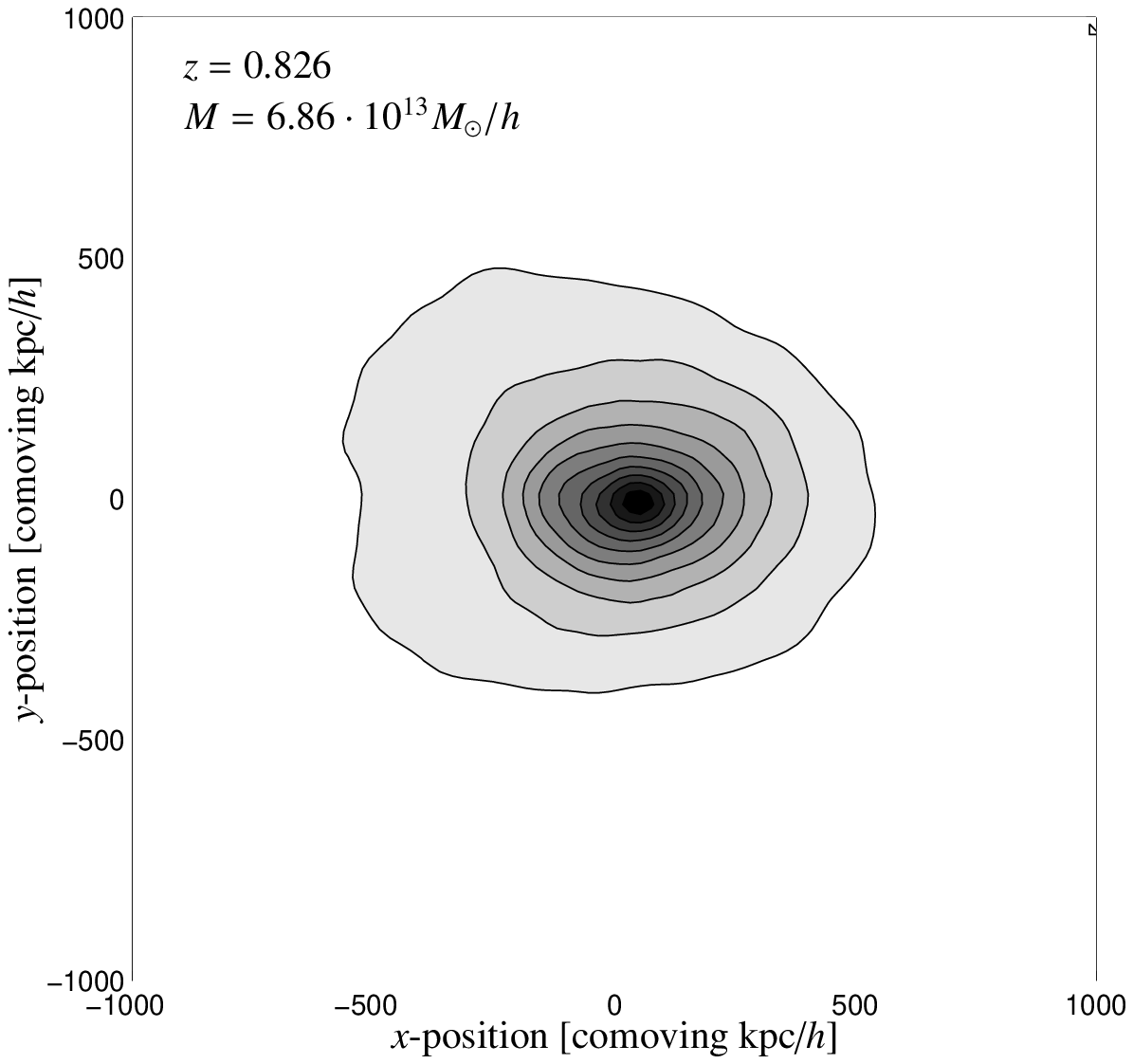}} &
\resizebox{5.5cm}{!}{\includegraphics{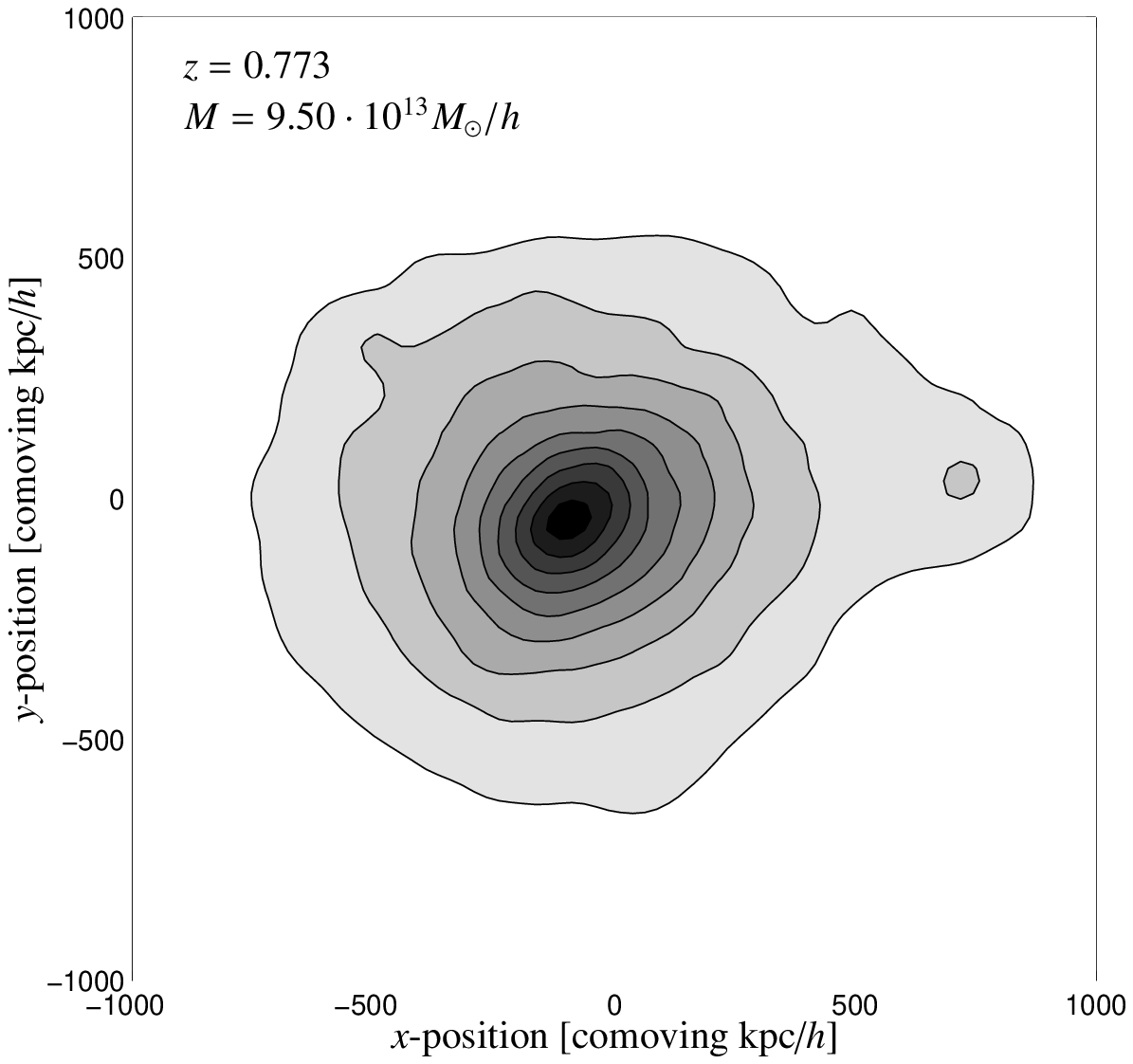}} \\
\resizebox{5.5cm}{!}{\includegraphics{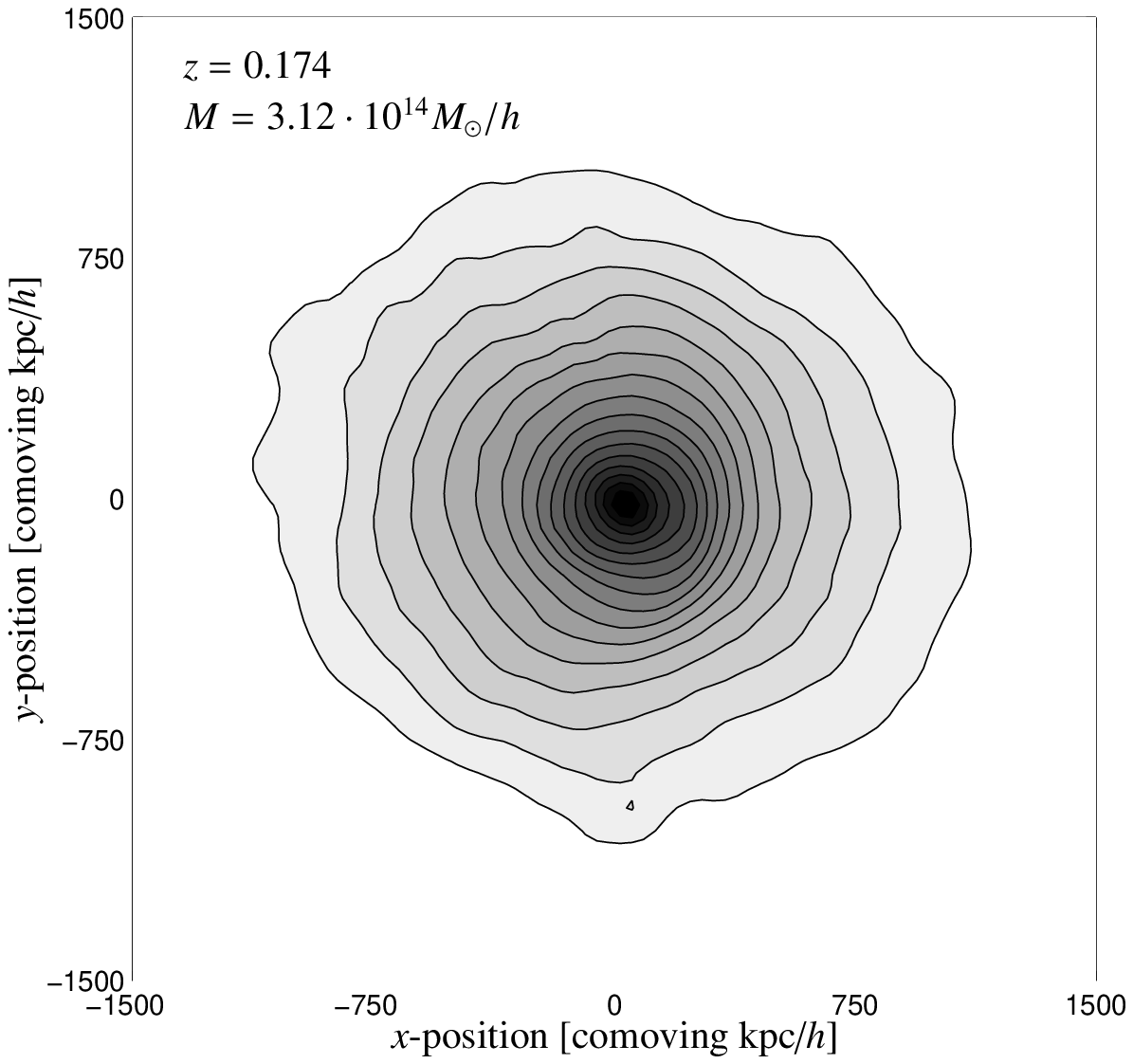}} &
\resizebox{5.5cm}{!}{\includegraphics{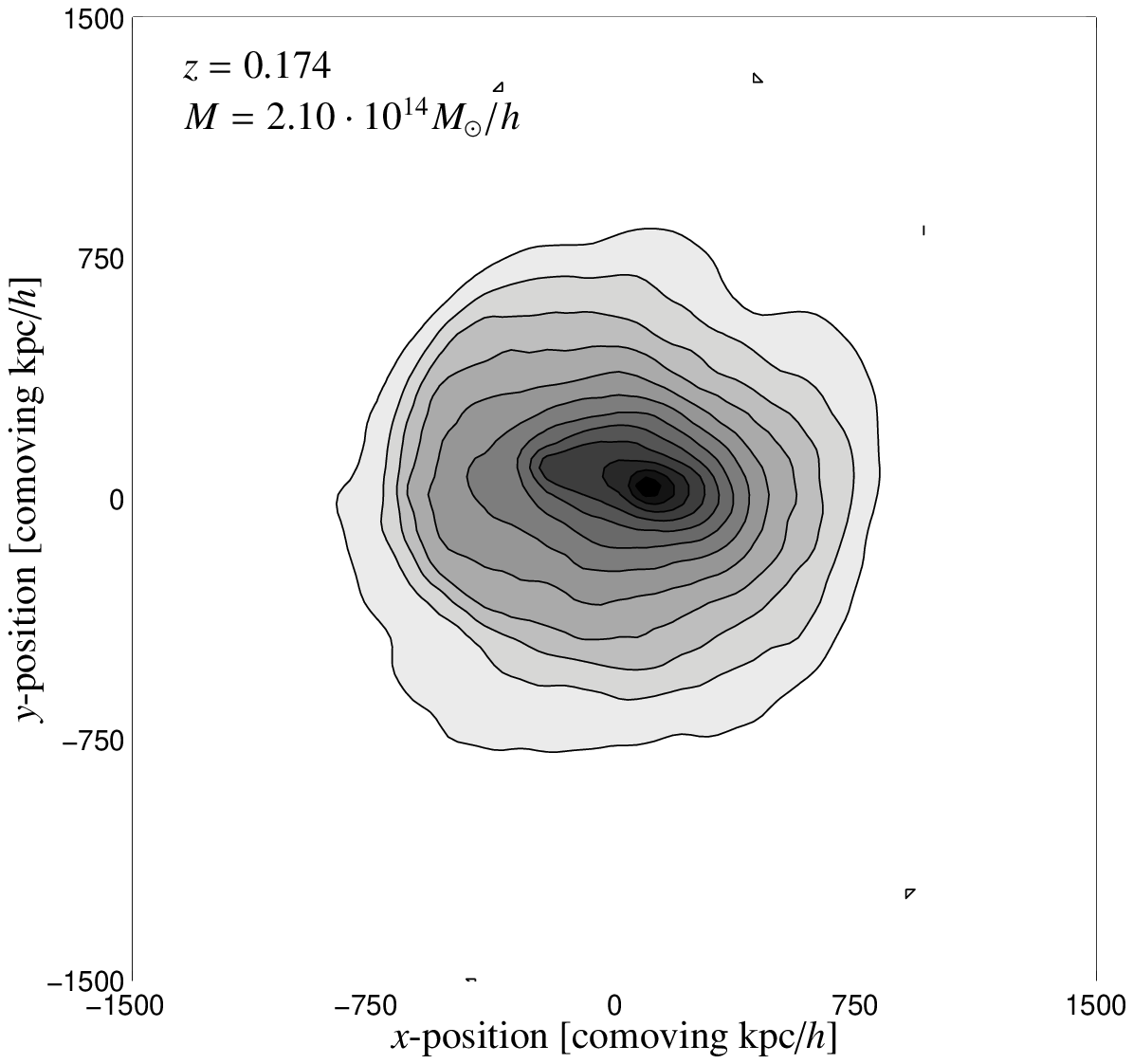}} &
\resizebox{5.5cm}{!}{\includegraphics{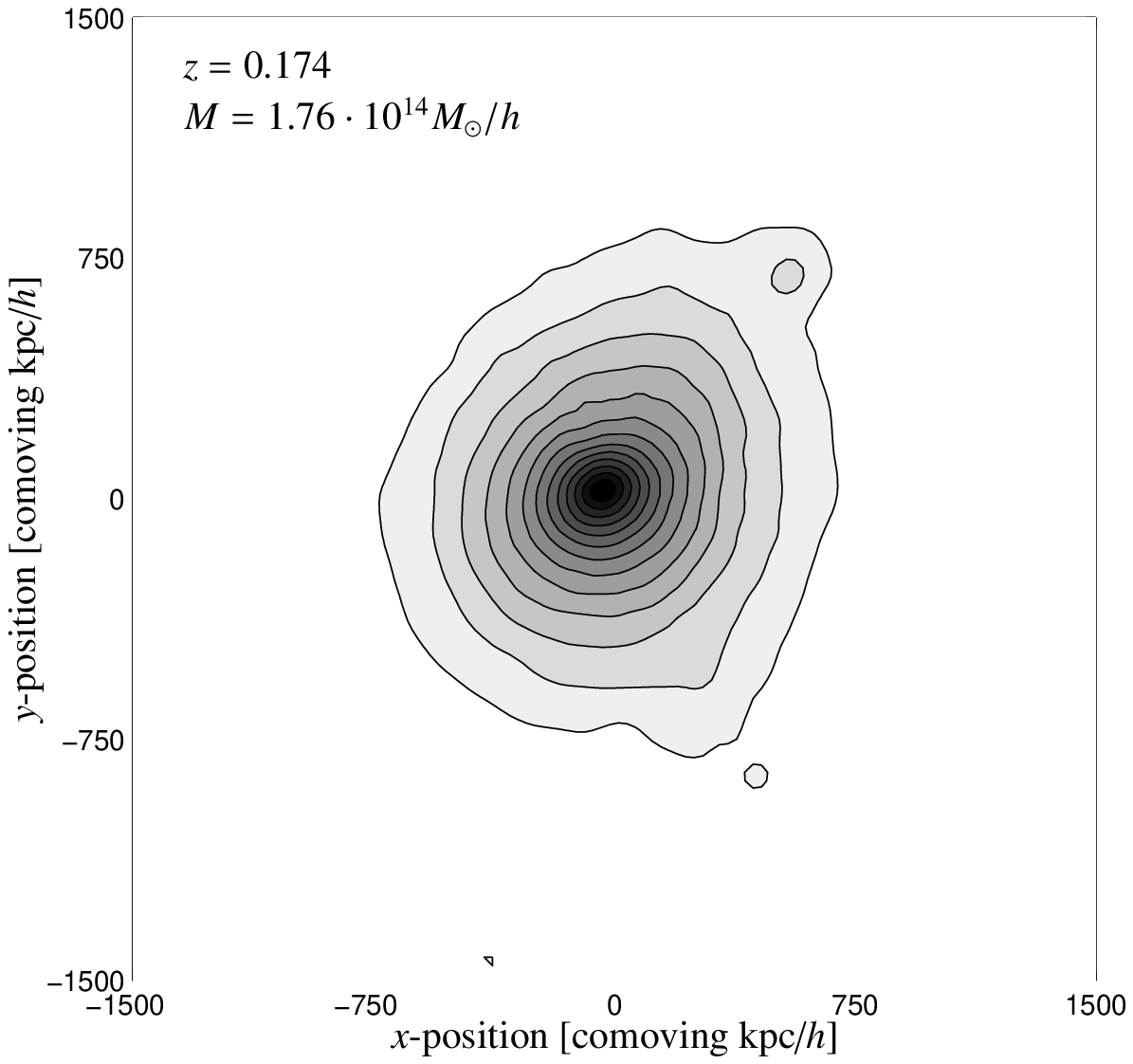}}
\end{tabular}
\caption{Picture book of Sunyaev-Zel'dovich clusters: The upper panel shows clusters at high redshifts of $z\simeq 0.8$,
in comparison to clusters at low redshifts of $z = 0.174$ in the lower panel. The columns contrast different
morphologies in an exemplary fashion: relaxed systems (left column), elongated clusters (centre column) and clusters in
the phase of minor merging or mass accretion (right column). The grey scale denotes the amplitude of
$y(\bmath{x})^\prime=\log\left[1+10^5\cdot y(\bmath{x})\right]$ and the contours have a logarithmically equidistant
spacing of $0.1\mbox{ dex}$, i.e. the lowest contour denotes a common value of $y=2.5\cdot 10^{-6}$.}
\label{fig_picturebook}
\end{figure*}

Square maps of the Compton-$y$ parameter of the selected clusters were generated by SPH projection of all member
gas particles onto a cubical grid with $128^2$ mesh points. The (comoving) side length $s$ of the map was adapted to
the cluster size, such that the comoving resolution $g=s/128$ of the grid is specific to a given map. Examples of
Sunyaev-Zel'dovich maps are given in Fig.~\ref{fig_picturebook}.

If the particle $p$ at position $\bmath{r}_p=\left(x_p,y_p,z_p\right)$ has a smoothing length $h_p$, an SPH electron
number density estimate $n_p$, and an SPH electron temperature $T_p$, the Compton-$y$ parameter at the pixel at position
$\bmath{x}$ is given by:
\begin{equation}
y(\bmath{x}) = \frac{\sigma_\mathrm{T} k}{m_e c^2}\frac{h_p^3}{g^2}
\sum_p\left[\:
\int\limits_{x-g/2}^{x+g/2}\!\!\!\dd x_p
\int\limits_{y-g/2}^{y+g/2}\!\!\!\dd y_p
\int\limits_{-h_p}^{h_p}\!\!\dd z_p\: w\left(\frac{r}{h_p}\right)\cdot n_p T_p\right]
\end{equation}

\begin{equation}
\mbox{with }r = \sqrt{(x_p-x)^2+(y_p-y)^2+z_p^2}\mbox{.}
\end{equation}

Here, we assumed complete ionisation and primordial element composition of the ICM for the determination of electron
number density and temperature. In this way we produced projections along each of the three coordinate axes. The
function $w$ is the spherically symmetric cubic spline kernel suggested by \citet{1985A&A...149..135M}, which was also
used in the SPH simulation:
\begin{equation}
w(u) = \frac{8}{\pi}\cdot
\left\{
\begin{array}
{l@{,\:}l}1-6u^2+6u^3 & 0 \leq u \leq 1/2\\2(1-u)^3 & 1/2 < u \leq 1 \\0 & u > 1
\end{array}
\right.
\mbox{ with } u = r / h_p\mbox{.}
\end{equation}

The fact that the kernel $w$ is defined on a compact support $u\in\left[0\ldots 1\right]$ greatly reduces the
computational effort.

\subsection{Cluster selection}\label{selection_criteria}
Clearly, the wavelet redshift estimation relies on the clusters not being in the state of violent merging. Apart from
the minimal mass of $M_\mathrm{min} = 5\cdot 10^{13} M_{\sun}/h$, that translates into a minimally required
line-of-sight Comptonisation amplitude $y_\mathrm{min}$, clusters have been selected in order to meet the following
prerequisites:

\begin{itemize}
\item{By visual inspection it was made sure that the clusters show but a single peak in the Compton-$y$ map in order to
exclude systems in the late phase of a merger.}

\item{The SZ morphology is required not to be too elongated. By fitting a 2-dimensional $\beta$-model
$y_\beta(\bmath{x})$ to the SZ profile $y_\mathrm{data}(\bmath{x})$, values for the semi-axes $r_x$ and $r_y$ are
derived. $90\%$ of the clusters within the selected sample have axis ratios $q = r_y/r_x$ smaller than 0.8 and
ellipticities $e=\sqrt{r_x^2-r_y^2} / r_x$ below 0.6.}

\item{Residual deviations from the canonical $\beta$-profile ought to be small. The $rms$-deviation $v$ of the cluster
from the best-fitting $\beta$-profile,
\begin{equation}
v = \sqrt{
\left\langle\left(\frac{y_\mathrm{data}(\bmath{x})-y_{\beta}(\bmath{x})}{y_{\beta}(\bmath{x})}\right)^2
\right\rangle_{\bmath{x}}}\mbox{,}
\end{equation}
was smaller than $25\%$ for $90\%$ of our clusters.}
\end{itemize}

\begin{figure}
\resizebox{\hsize}{!}{\includegraphics{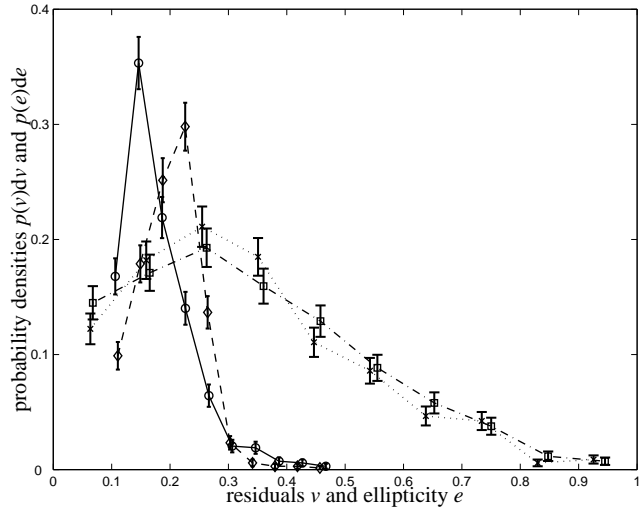}}
\caption{Selection criteria: distribution of residual deviations $v$ from the best-fitting $\beta$-profile for
unsmoothed (solid, circles) and smoothed (dashed, diamonds) maps. The second set of lines shows the distribution of the
ellipticies $e$ with (dash-dotted, squares) and without smoothing (dotted, crosses).}
\label{fig_selection}
\end{figure}

Applying these selection criteria, 10 clusters were selected from each of the 23 redshift bins, yielding with the three
orthogonal projections of each cluster a total number of 690 maps with which the accuracy of the wavelet method in
estimating redshifts was assessed. The distributions of the ellipticities $e$ and the integrated residuals $v$ are shown
in Fig.~\ref{fig_selection}. The same distributions were derived for the smoothed cluster maps, were the effects of
finite instrumental resolution have been incorporated. As Fig.~\ref{fig_selection} suggests, the beam does not have a
major impact on the morphological properties of the cluster sample, which is due to its narrowness of only $1\arcmin$
(FWHM).

\subsection{CMB map generation}\label{cmb_fast_descr}
CMB anisotropies are assumed to be a particular realisation of a {\em Gaussian random field}. Aiming at simulating a
realisation of the CMB on a square, flat map, we take temperature fluctuations $\theta(\bmath{\phi})$ relative to the
average CMB temperature of $\bra T\ket = 2.726\mathrm{ K}$ to be the independent random field,
\begin{equation}
\theta(\bmath{\phi})\equiv\frac{T(\bmath{\phi})-\langle T\rangle}{\langle T\rangle}\mbox{.}
\label{random field}
\end{equation}

The flat, two-dimensional power spectrum $P_\theta(\ell)$ is defined via:
\begin{equation}
\left\langle \Theta(\bmath{\ell})\Theta^*(\bmath{\ell}')\right\rangle
\equiv(2\pi)^{-2}\delta_D(\bmath{\ell}-\bmath{\ell}')P_\theta(|\bmath{\ell}|)\mbox{,}
\label{simulating CMB power spectrum}
\end{equation}
where $\Theta(\bmath{\ell})$ denotes the Fourier transform of $\theta(\bmath{\phi})$. The simulation of the CMB
temperature fluctuations on a flat square map now consists of the following two steps:

\begin{itemize}
\item{The angular power spectrum $C_\ell$ is computed for the flat \mbox{$\Lambda$CDM}-Universe using the {\tt CMBfast}
code by \citet{1996ApJ...469..437S}. In addition to the cosmological parameters being already described in
Sect.~\ref{sim}, we use adiabatic initial conditions and set the primordial He-mass fraction to $X_\mathrm{He} = 0.24$
and the Thomson optical depth to $\tau = 0.17$ \citep{2003astro.ph..2209S}. The angular power spectrum of the CMB is
normalised to COBE data. Since the SZ effect distorts the CMB only on small angular scales, the flat sky approximation
$\ell\gg 1$ is fulfilled and it is appropriate to replace the spherical harmonics with plane waves.
\citet{2000PhRvD..62d3007H} has shown that the 2-dimensional flat power spectrum $P_\theta(\ell)$ is approximately equal
to its angular analogue: $C_\ell\simeq P_\theta(\ell)$.}

\item{Then, Gaussian random variables are generated on a complex two-dimensional grid in Fourier space with variance
$\sigma^2(\ell)= P_\theta(\ell)$ according to the absolute value of their wave vectors $\bmath{\ell}$. Inverse Fourier
transform yields a realisation of the temperature anisotropies $\theta(\bmath{\phi})$.}
\end{itemize}

\subsection{Simulated single-frequency SZ observations}\label{sim_obs_descr}
For SZ clusters observed with a single-frequency interferometer \citep[e.g., the CBI
experiment,][]{2002ApJ...568...38H}\footnote[1]{\tt http://www.astro.caltech.edu/\textasciitilde tjp/CBI/}, it is
important to examine the applicability of the $X_q(\sigma)$-statistic. For the purpose of this paper, it suffices to
consider observations at small frequencies $\nu$. Thus, the Compton-$y$ maps are combined with realisations of the CMB
fluctuations by taking advantage of eqn.~(\ref{sz_temp_decr}) in the Rayleigh-Jeans limit,
\begin{equation}
T(\bmath{\phi}) = \left[1 - 2 y(\bmath{\phi})\right] \left[1+\theta(\bmath{\phi})\right]
\bra T\ket\mbox{.}
\end{equation}

\begin{figure}
\resizebox{\hsize}{!}{\includegraphics{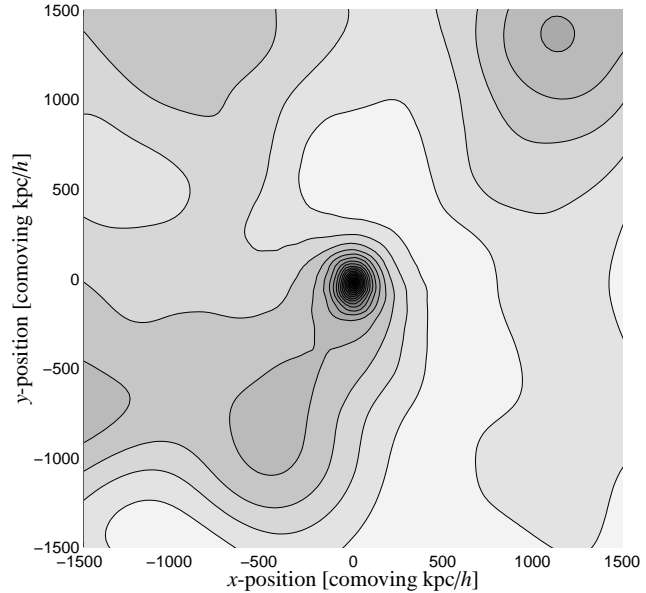}}
\caption{Simulated temperature map of the CMB combined with a foreground SZ cluster at $z=0.102$ with virial quantities
$M_\mathrm{vir} =2.2\cdot10^{14}M_{\sun}/h$, $r_\mathrm{vir} = 1.47\mbox{ Mpc}/h$ and $kT_\mathrm{vir} = 1.52\mbox{
keV}$. At the cluster centre, the SZ temperature decrement amounts to $-1.8\mbox{ mK}$ and the CMB temperature
fluctuation with the highest amplitude is equal to $0.23\mbox{ mK}$. A total of 30 linearly spaced isothermals are
drawn. In this case, the comoving scale 1~Mpc/$h$ corresponds to $11.5\arcmin$.}
\label{fig_cmbmap}
\end{figure}

Fig.~\ref{fig_cmbmap} shows the Compton-$y$ map of a nearby cluster of $2.2\cdot 10^{14} M_{\sun}/h$ at redshift $z
=0.102$ combined with a patch of the CMB constructed by the algorithm described above. In this map, the average CMB
temperature $\bra T\ket$ was subtracted. In order tomimic observations, the resulting combined maps are smoothed with a
Gaussian beam with FWHM of $\sqrt{8\ln(2)}\cdot\sigma_b =1\arcmin$.

In the case of multi-frequency SZ observations the SZ signature can be easily distinguished from the CMB signal.
Therefore, for these cases the CMB background is ignored and not included in the simulated cluster SZ images.
Nevertheless, finite instrumental resolution was taken care of and the SZ maps were convolved with a Gaussian kernel of
$\sqrt{8\ln(2)}\cdot\sigma_b = 1\arcmin$ (FWHM).

\section{Analysis}\label{ana}
In this section, the analysis is explained step by step: After introducing the wavelet families
(Sect.~\ref{ana_wavelet_basis}), the wavelet spectrum and the parameters deduced from it are described
(Sect.~\ref{ana_wavelet_measurement} and Sect.~\ref{ana_wavelet_spec}). The correlations of the wavelet spectral
parameters with physical quantities are discussed (Sect.~\ref{ana_correlation}). The measurement principle and the
breaking of degeneracy is illustrated in Sect.~\ref{ana_yardstick}. Successively, the intercorrelation of the wavelet
parameters and the shape of the parameter space is explored by principal component analysis (Sect.~\ref{ana_princomp}).
Then, gauge functions for modelling the redshift dependence ofthe parameters are proposed (Sect.~\ref{ana_zestimate}).
An important issue for single-frequency interferometers is the influence of primary CMB fluctuations on the wavelet
spectrum and their suppression (Sect.~\ref{ana_cmb_infl}). Finally, the redshift of the clusters are estimated by
maximum likelihood techniques (Sect.~\ref{ana_accuracy}).

\subsection{Wavelet basis functions}\label{ana_wavelet_basis}
\begin{figure}
\resizebox{\hsize}{!}{\includegraphics{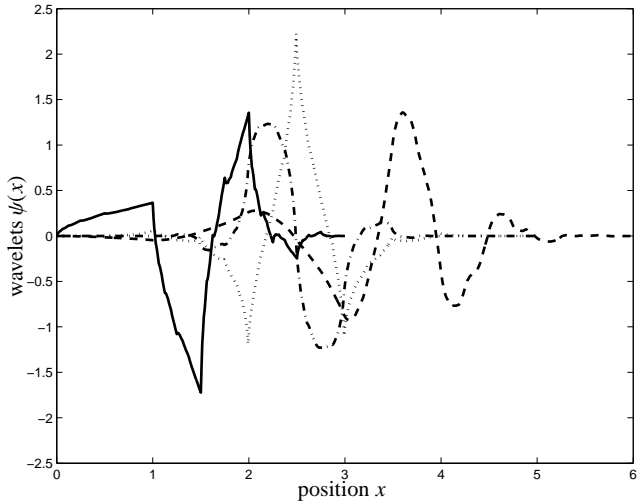}}
\caption{The wavelet basis functions $\psi(x)$ chosen for the analysis: symlet {\em sym2} (solid), Daubechies' wavelet
{\em db4} (dashed), coiflet {\em coif1} (dotted) and the biorthogonal wavelet {\em bior1.3} (dash-dotted).}
\label{fig_wavelet_base}
\end{figure}

In the analysis a wide range of wavelets with different functional shapes was employed, although the {\em symlet}
wavelet basis introduced by \citet{1993ASAJ...93.1671D} yielded particularly good results. Due to their symmetry and
peakiness, {\em symlets} are seemingly especially suited for analysing SZ morphologies, because spiky features are
useful for edge detection, and since apparent cluster size is the primary observable, the usage of a peaked wavelet is
justified. Other wavelet families that found application in our analysis were Daubechies' wavelets, coiflets and
biorthogonal wavelets. Fig.~\ref{fig_wavelet_base} compares the functional shape of the different wavelet families.

The analysis proceeds by measuring the wavelet moments on smoothed comoving maps of the Compton-$y$ parameter following
the definition in Sect.~\ref{def_wavelet}. The scale $\sigma$ of the resulting wavelet spectrum is then converted to
angular units. Because our SZ maps are computed on a grid of $128^2$ mesh points with adaptively chosen side length for
each cluster, our dynamical range of the wavelet spectra always comprises approximately two decades. However, this is
no fundamental limitation of this approach because the maps are featureless below the smoothing scale of
$1\arcmin$ (FWHM).

\subsection{Measurement of wavelet quantities}\label{ana_wavelet_measurement}
In order to derive the actual flux decrement or, equivalently, the decrement in antenna temperature from the
line-of-sight Compton-$y$ amplitude, its value at each pixel needs to be multiplied with the solid angle it subtends.
For the conversion, a standard \mbox{$\Lambda$CDM}-cosmology was assumed, the parameters of which have already been
given in Sect.~\ref{sim}. Thus, the pixel amplitudes were modified according to:
\begin{equation}
y(\bmath{x})\longrightarrow y(\bmath{\phi})=y(\bmath{x})\cdot4\arctan^2\left[\frac{g}{2 w(z)}\right]\mbox{,}
\end{equation}
where $w(z)$ is the comoving distance in the model cosmology and $g$ denotes the comoving size of a single pixel. It
should be emphasised that the wavelet coefficients $\chi(\bmath{\mu},\sigma)$ are evaluated on a comoving grid, which
has been adapted to the cluster size before converting the wavelet scale $\sigma$ to angular units. This, however,
should not pose a problem for real observations, provided the sampling scale is of the same order of magnitude as the
angular scale of our finest pixels.

In order to obtain dimensionless quantities, the unit of the wavelet $\psi_\sigma(\bmath{x})$ has been set to
inverse steradians, such that the wavelet expansion coefficients $\chi(\bmath{\mu},\sigma)$ and the wavelet moments
$X_q(\sigma)$ are dimensionless, irrespective of $q$. For numerical convenience, the pixel amplitudes in the combined SZ
maps have been multiplied with $10^{12}$.

The summation in the definition of the wavelet moment $X_q(\sigma)$ in eqn.~(\ref{wavelet_spec}) discards the
information about the position $\bmath{\mu}$ at which the wavelet expansion coefficient $\chi(\bmath{\mu},\sigma)$ is
evaluated. Consequently, the position of a cluster inside the observing frame does not influence the wavelet
decomposition.

\subsection{Wavelet spectrum of SZ cluster maps}\label{ana_wavelet_spec}
\begin{figure}
\resizebox{\hsize}{!}{\includegraphics{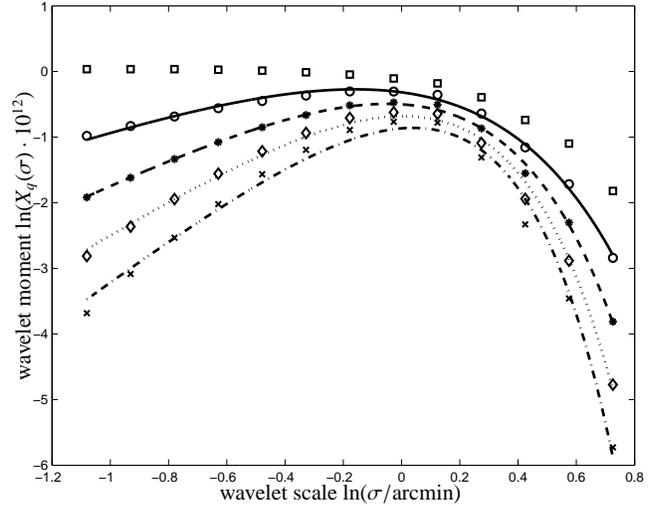}}
\caption{The spectrum of wavelet moments, together with the fitting formula~(\ref{fitting_formula}) for increasing
wavelet moment order $q$: $q = 2$ (squares), $q =3$ (circles, solid), $q = 4$ (stars, dashed), $q = 5$ (diamonds,
dotted) and $q=6$ (crosses, dash-dotted) for a single cluster. The wavelet moments $X_q(\sigma)$ followed from wavelet
expansion with the {\em sym2}-wavelet.}
\label{fig_wavelet_spec}
\end{figure}

Due to the lack of any analytical generalisation of eqn.~(\ref{analytic_spec}) for $q\neq 2$, deviations of the
Compton-$y$ map from a King profile and wavelets other than the simple Mexican hat, we decided to explore
phenomenological functions for describing the wavelet spectrum. The simplicity of the shape of the wavelet spectrum
shown in Fig.~\ref{fig_wavelet_spec} implies that the model function,
\begin{equation}
\ln X_q(\sigma) \simeq a +s\ln\left(\sigma/\sigma_0\right) - \sigma/c\mbox{,}\label{fitting_formula}
\end{equation}
is able to extract all apparently contained information, i.e. the spectrum is described by means of three quantities:
the amplitude $a$, the slope $s$ and a break at $c$. The parameter $\sigma_0$ has been included
in eqn.~(\ref{fitting_formula}) in order to obtain a formula that is dimensionally correct, although it does not yield
any new information and this specific degree of freedom is already described by the variable $a$.

The usage of eqn.~(\ref{fitting_formula}) implicitly discards information about asphericity and effectively determines
an average of the cluster's extension along its major axes. The problem would be significantly complicated by including
asymmetry and considering vectorial nature of $\sigma$ \citep[see][]{1998ApJ...500L..87Z,2001ApJ...561..600Z}.

Because the cutoff parameter $c$ is of great importance to our analysis, it needs to be derived reliably. Thus, the
order of wavelet moments $q$ was restricted to $q\geq 3$, because larger $q$-values facilitate the determination of $c$.
From Fig.~\ref{fig_wavelet_spec} it is obvious that an increase in $q$ suppresses the value of $X_q(\sigma)$ at small
scales $\sigma$ such that the curve develops a maximum in the vicinity of $c$. Additionally, by the choice of large
values for $q$, the wavelet expansion coefficients $\chi(\mu,\sigma)$ dominated by CMB noise are suppressed relative to
those obtained in the central part of the cluster and consequently higher order wavelet moments $X_q(\sigma)$ provide a
cleaner measurement. The range of sensible $q$-values is restricted by the fact that for increasing $q$ the moment
$X_q(\sigma)$ is successively dominated by the largest wavelet expansion coefficient $\chi(\mu,\sigma)$ and does no
longer contain information of the structure to be analysed. In order to stabilise the fitting procedure we interpolate
in between the wavelet moments $X_q(\sigma)$. This is justified because we expect a smooth variation of the wavelet
spectrum according to Sect.~\ref{application}.

\subsection{Correlations with physical quantities}\label{ana_correlation}
The parameters derived from the fit to the spectrum of wavelet coefficients have a physical interpretation. As
illustrated in Sect.~\ref{application}, the wavelet spectrum breaks at the cluster scale.

\begin{figure}
\resizebox{\hsize}{!}{\includegraphics{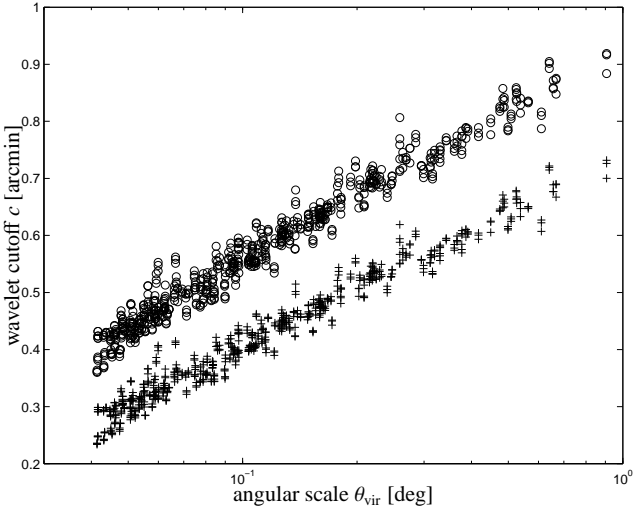}}
\caption{Wavelet measured cluster size $c$ versus angular extension $\theta_\mathrm{vir}$ for increasing wavelet moment
order $q$: $q = 3$ (circles) and $q = 6$ (crosses) without including CMB fluctuations. The $c$-values have been
determinedwith the {\em sym2}-wavelet.}
\label{fig_ctheta_correlation}
\end{figure}

Therefore, one expects a correlation between the angular size of the cluster $\theta_\mathrm{vir}$ and the {\bf cutoff
$c$}, as shown by Fig.~\ref{fig_ctheta_correlation}. Increasing weighting exponents $q$ shift the regression line to
smaller values of $c$, which can be understood by the fact that larger values of $q$ suppress small wavelet expansion
coefficients arising at the outskirts of the cluster, which in turn leads to a break in the wavelet spectrum at smaller
scales.

\begin{figure}
\resizebox{\hsize}{!}{\includegraphics{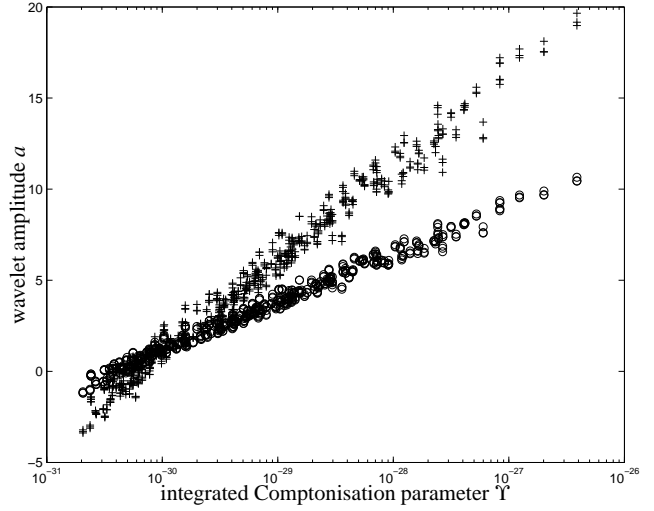}}
\caption{Wavelet amplitude $a$ as a function of integrated Comptonisation parameter $\Upsilon$ for different weighting
exponents $q$: $q =3$ (circles) and $q= 6$ (crosses), again without taking CMB fluctuations into account. As the
analysing wavelet, the {\em sym2}-wavelet was chosen.}
\label{fig_ay_correlation}
\end{figure}

Similarly, the {\bf amplitude $a$} determined by the fit is proportional to the integrated Compton-$y$ flux,
\begin{equation}
\Upsilon = \int\dd^2\phi\: y(\bmath{\phi}) =
\frac{kT_\mathrm{vir}}{m_e c^2}\frac{\sigma_T}{\dang(z)^2}\frac{1+f_H}{2} f_B
\frac{M_\mathrm{vir}}{m_\mathrm{p}}\mbox{,}
\label{eqn_szflux}
\end{equation}
as illustrated by Fig.~\ref{fig_ay_correlation}. Here, $f_b$ denotes the baryon fraction, $f_H$ the hydrogen fraction,
which determines the elemental composition and has been set to the primordial value of 0.76, and $m_\mathrm{p}$ is the
proton mass. $\dang(z)$ is the angular diameter distance in our cosmology.

The normalisation $a$ of the wavelet moments $X_q(\sigma)$ shows a steeper dependence on the integrated Comptonisation
parameter $\Upsilon$ for larger choices of $q$, which is explained by the following argument: The amplitude $a(q)$
reflects the normalisation of the wavelet moments $X_q(\sigma)$. The integral in eqn.~(\ref{wavelet_spec}) is dominated
by the largest wavelet expansion coefficient $\chi(\bmath{\mu},\sigma)$, taken to the $q^\mathrm{th}$ power. On the
other hand, the wavelet expansion coefficients $\chi(\bmath{\mu},\sigma)$ are proportional to integrated Comptonisation
$\Upsilon$, resulting in observed relation $\ln\left[X_q(\sigma)\right]\propto a\propto q\cdot\ln(\Upsilon)$.

The {\bf slope $s$} is a measure of instrumental smoothing: Placing the same cluster at different redshifts would result
in a blurred image of the more distant one. Keeping in mind that there is a close analogy between wavelet- and
Fourier-transforms (as explained in Sect.~\ref{analogy}), the wavelet moment $X_q(\sigma)$ as a function of $\sigma$ can
be interpreted as the variance of the wavelet-filtered field. The instrumental beam introduces an additional filtering
to the Compton-$y$ map (compare Sect.~\ref{finite_resolution}) and would cause the Fourier spectrum to drop at smaller
values of the wave vector $k$, because the instrumental beam constitutes effectively a low-pass filter that is erasing
structures smaller than its extension. Comparing clusters at different redshifts, it is clear that the drop in power
happens at smaller scales in the case of the more distant cluster. Then the slope $s$, defined as
$\dd\ln X_q(\sigma)/\dd\ln\sigma$ for $\sigma\ll r_c$, is larger in the case of a resolved cluster compared to an
unresolved cluster. This measure of the influence of finite instrumental smoothing varies only by a factor of two in the
redshift and mass range considered here, but nevertheless serves as an indicator of cluster distance.

\subsection{Measurement principle}\label{ana_yardstick}
Now, it is necessary to illustrate how a measurement of the total Comptonisation $\Upsilon$ and of the angular size
$\theta_\mathrm{vir}$ suffices to derive a distance estimate. For that purpose, clusters are placed at unit distance
and the distance dependences of the wavelet amplitude $a$ and the the cutoff $c$ are removed by the following formulae,
since $a$ is a logarithmic measure of flux inside an solid angle element $\Upsilon$ and $c$ is a logarithmic measure of
of angular extension $\theta_\mathrm{vir}$:
\begin{eqnarray}
a_0 & = & a(z) + 2\cdot\ln\left( d_a(z)\right) \\
c_0 & = & c(z) + \ln\left( d_a(z)\right)\mbox{.}
\label{unit_distance}
\end{eqnarray}

By applying simple scaling arguments, one expects the ratio $a_0 / c_0$ to be equal to 5: From the wavelet amplitude
$a$ one obtains
$a_0\propto\ln\left(\Upsilon\cdot d_A(z)^2 \right)\propto\ln\left(M_\mathrm{vir}\cdot T_\mathrm{vir}\right)$.
Furthermore, from the spherical collapse model follows, that $T_\mathrm{vir}\propto M_\mathrm{vir}^{2/3}$
\citep{1995MNRAS.275..720N}, which yields, together with $M_\mathrm{vir}\propto r_\mathrm{vir}^3$, the relation
$a_0\propto\ln\left(r_\mathrm{vir}^5\right)$. Substituting $c_0\propto\ln\left(r_\mathrm{vir}\right)$ gives the
final result $a_0 / c_0 = 5$.

\begin{figure}
\resizebox{\hsize}{!}{\includegraphics{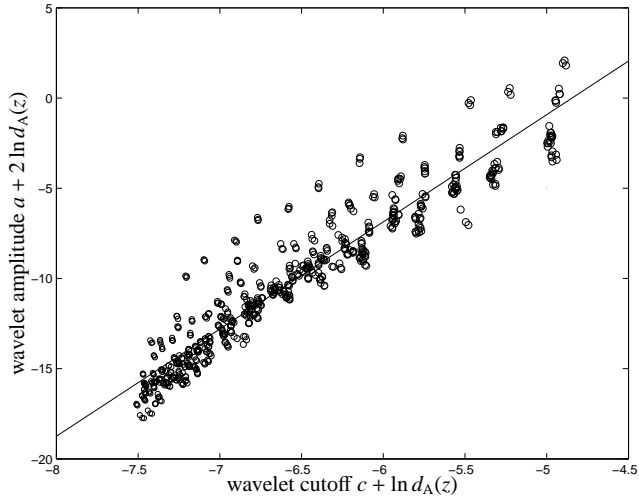}}
\caption{Distance corrected wavelet amplitude $a_0 = a(z) + 2\cdot\ln\left( d_a(z)\right)$ as a function of distance
corrected wavelet cutoff parameter $c_0 = c(z) + \ln\left( d_a(z)\right)$. The values have been determined in fits to
the wavelet spectrum $X_3(\sigma)$, that has been derived with the {\em sym2}-wavelet as the analysing wavelet.}
\label{fig_unit_distance}
\end{figure}

Fig.~\ref{fig_unit_distance} nicely illustrates how the degeneracy is broken and how a simple measurement of flux and
angular extension suffices in order to derive a distance estimate: A crude fit to the distance corrected wavelet
amplitude $a_0$ as a function of distance corrected wavelet cutoff parameter $c_0$ yields a slope of approximately 5.8,
which corresponds well to the slope of $\sim 5$ expected from the theoretical consideration outlined above. If,
hypothetically, the ratio $a_0 / c_0$ was equal to 2, the measurements of flux and angular size would be completely
degenerate and would not yield any distance information. This case corresponds to disks of equal surface brightness,
where measurements of flux and angular size are completely degenerate and do not yield any distance information at all.
It should be noted, that by adopting the usual scaling relations, one introduces a systematic error in slope that can
amount to $\simeq 20\mbox{\%}$ error.

\subsection{Principal component analysis}\label{ana_princomp}
In order to rate the extent to which the parameters derived in the fit to the spectrum of wavelet moments
are independent, a principal component analysis \citep[PCA, see e.g.][]{1964MNRAS.127..493D} was performed. The
PCA is determining a transformation to a new orthogonal coordinate system in parameter space spanned by $a$, $c$
and $s$, such that the variance along the first axis is maximised.

The first eigenvector of the matrix that describes the change of basis by the PCA reads $x_\mathrm{PCA} =
(0.65, 0.70, 0.32)$ which has been derived for the spectral parameters for $q=3$ and with the {\em sym2}-wavelet as the
analysing wavelet. The values similar in magnitude suggest that the variation in the data set is contained in all three
parameters $a$, $c$ and $s$.

\begin{table}
\vspace{-0.1cm}
\begin{center}
\begin{tabular}{rrrrr}
\hline\hline
\vphantom{\Large A}%
					& $q=3$ 	& $q=4$ 	& $q=5$ 	& $q=6$		\\
\hline
\vphantom{\Large A}%
$1^\mathrm{st}$ principal component 	& 95.6\% 	& 94.2\% 	& 92.8\% 	& 91.5\%	\\
$2^\mathrm{nd}$ principal component 	& 2.7\% 	&  4.2\% 	&  5.5\% 	& 6.7\%		\\
\hline
\end{tabular}
\end{center}
\caption{Result of the PCA. The variance explained by the first and second principal component as a function of wavelet
order $q$. Here, no CMB fluctuations were included.}
\label{table_pca}
\end{table}

As can be read off from table~\ref{table_pca}, the parameter space is tightly constrained and all three parameters are
interrelated, such that the data points form a narrow ray in parameter space. This result holds irrespective of the
choice of $q$, although the scatter increases with higher choices for $q$. The values in table~\ref{table_pca} have been
determined without considering CMB fluctuations. Given the physical interpretation of the wavelet amplitude $a$ and the
cutoff $c$, it is obvious that the tight correlation can be traced back to the self-similarity of clusters and the
cluster scaling relations linking $T_\mathrm{vir}$, $M_\mathrm{vir}$ and $r_\mathrm{vir}$ that follow from the
spherical collapse model. The scaling relations for SZ quantities derived by \citet{2003astro.ph..8074D} support
this view. This shows together with Sect.~\ref{ana_correlation} and Sect.~\ref{ana_yardstick} hat both the cutoff $c$
and the amplitude $a$ are functions of $M_\mathrm{vir}$. By only considering systems close to virialisation we thus
expect a fundamental narrow ray in parameter space which can be described by a single principal component.

\subsection{Redshift dependence of the wavelet parameters}\label{ana_zestimate}
\begin{figure}
\resizebox{\hsize}{!}{\includegraphics{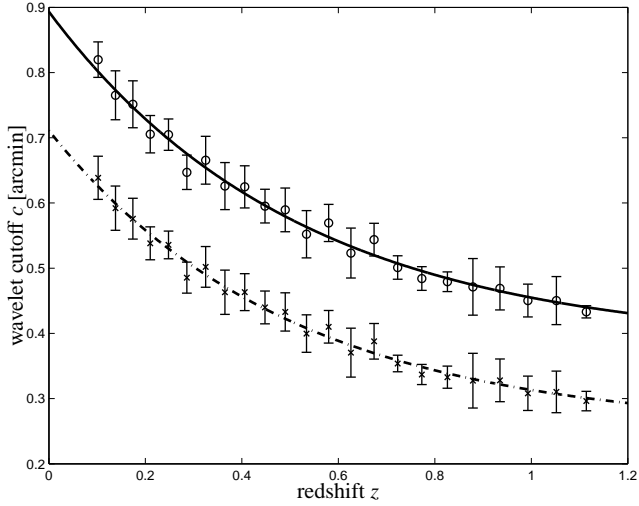}}
\caption{Dependence of the wavelet cutoff parameter $c$ on redshift $z$ without considering CMB fluctuations for $q=3$
(circles, solid) and $q=6$ (crosses, dash-dotted). The analysing wavelet was the {\em sym2}-wavelet.}
\label{fig_cutoff_decrease}
\end{figure}

\begin{figure}
\resizebox{\hsize}{!}{\includegraphics{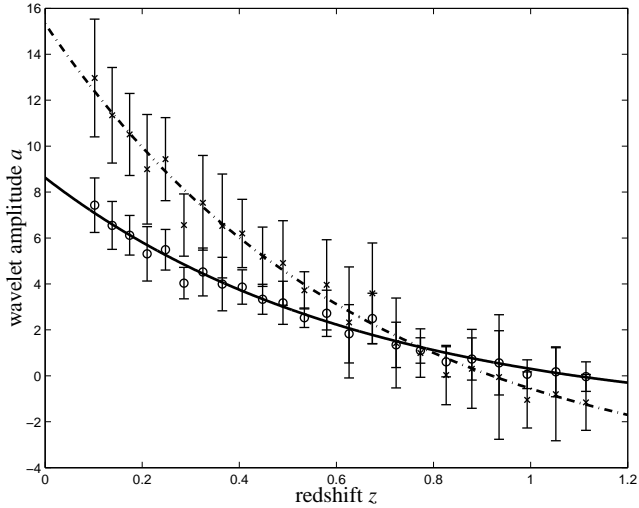}}
\caption{Dependence of the wavelet amplitude parameter $a$ on redshift $z$ without including CMB fluctuations for $q=3$
(circles, solid) and $q=6$ (crosses, dash-dotted). The quantities have been determined with the {\em sym2}-wavelet.}
\label{fig_amplitude_decrease}
\end{figure}

The parameters $a$ and $c$ are expected to decrease with increasing redshift $z$, the reason for which is quite
apparent: The angular diameter $\theta_\mathrm{vir}$ and the integrated Comptonisation $\Upsilon$ decreases because of
the increasing angular diameter distance $\dang(z)$ that enters $\theta_\mathrm{vir}$ linearly and $\Upsilon$
quadratically. Furthermore, clusters accrete matter during their formation history and thus are on average more massive
at later times, i.e. at smaller redshifts $z$
\citep[see, e.g.,][]{2002ApJ...568...52W,2002MNRAS.331...98V,2003MNRAS.339...12Z}. From the physical point of view, the
dependence of $a$ and $c$ on redshift $z$ is far from trivial, and therefore, their functional behaviour is described by
an empirical approach. Among others, the exponential function provides a good fit to the data, as illustrated by
Figs.~\ref{fig_cutoff_decrease} and \ref{fig_amplitude_decrease}:
\begin{equation}
\label{wavelet_par}
\label{eqn_gauge}
x(z) = x_1\exp\left(-\frac{z}{x_2}\right) + x_3\mbox{, where }x\in\left\{a,c,s\right\}\mbox{.}
\end{equation}

The optimised parameters $x_i$, $i\in\left\{1,2,3\right\}$, for $x\in\left\{a,c,s\right\}$ in the gauge functions
eqn.~(\ref{eqn_gauge}) are given in table~\ref{table_gauge_parameters} for the case $q=3$. It should be emphasised that
the parameters stated are only valid for image analysis with the {\em sym2}-wavelet, where the maps have been smoothed
with a Gaussian kernel with $1\arcmin$ (FWHM) and the considered cluster sample, which is defined by the selection
criteria laid down in Sect.~\ref{selection_criteria} and the minimal mass of $5\cdot 10^{13}M_{\sun}/h$.

\begin{table}
\vspace{-0.1 cm}
\begin{center}
\begin{tabular}{llccc}
\hline\hline
\vphantom{\Large A}%
parameter	& variable 		& $i=1$ 	& $i=2$ 	& $i=3$ 	\\
\hline \vphantom{\Large A}%
amplitude 	& $a$			& 10.5837	& 0.6475	& -1.9570	\\
cutoff [arcmin]	& $c$			&  0.5124	& 0.5165	& 0.3809	\\
slope  		& $s$			&  1.3423	& 0.4144    	& 1.3803	\\
\hline
\end{tabular}
\end{center}
\caption{Fitting values for the gauge functions defined in eqn.~(\ref{eqn_gauge}) for the cluster sample at hand and the
{\em sym2}-wavelet basis. The order of the wavelet moment $X_q(\sigma)$ has been set to $q=3$. The values have been
derived without taking CMB fluctuations into account.}\label{table_gauge_parameters}
\end{table}

\subsection{Influence of CMB noise}
\label{ana_cmb_infl}
\begin{figure}
\resizebox{\hsize}{!}{\includegraphics{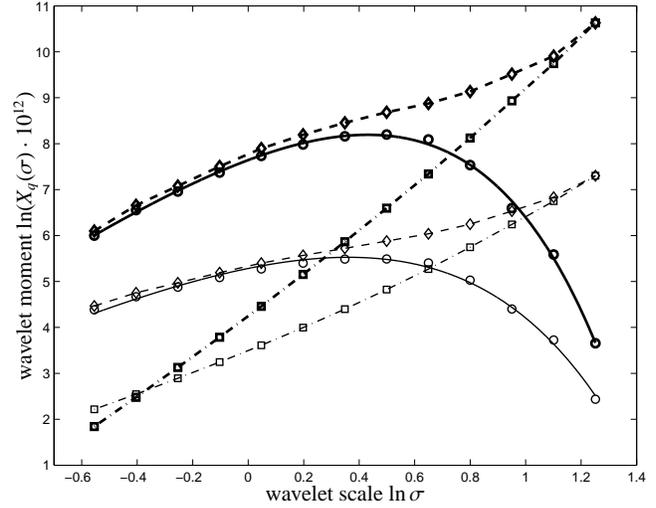}}
\caption{Changes to the wavelet spectrum of a single cluster caused by the fluctuating CMB: unperturbed wavelet spectrum
of the SZ cluster (circles, solid), of the pure CMB (squares, dash-dotted) and of the combined map (diamonds, dashed).
Data points were derived from simulated data and the joining line in the case of the unperturbed wavelet spectrum the
result of the fitting functions described by eqn.~(\ref{fitting_formula}). The order of the wavelet moment is $q=6$
(thick) and $q=4$ (thin). Again, the analysing wavelet is the {\em sym2}-wavelet.}
\label{fig_cmb_wavelet_spectrum}
\end{figure}

Clusters at high redshift $z$ are characterised by their small angular scale on which the underlying CMB is represented
by a smooth gradient due to Silk damping \citep{1968ApJ...151..459S}. In this case the wavelet analysis produces the
same results irrespective of the CMB noise owing to the distinct morphological feature of the cluster on top of the
smooth CMB gradient. Once clusters at lower redshifts reach angular sizes comparable to characteristic scales of CMB
fluctuations, the wavelet analysis has to be made more sophisticated.  This complication in the wavelet analysis arises
because wavelets are primarily suited for determining morphological features rather than solely singling out high
amplitude characteristics. Because the angular scale of the clusters ranges between $10\arcmin$ and $1\arcmin$, which
corresponds to multipole orders of $\ell\simeq10^3\ldots10^4$, it suffices to consider the Silk damping tail of the
angular power spectrum of the CMB. In the wavelet spectrum $X_q(\sigma)$ this translates into an additional approximate
power-law component $X^\mathrm{CMB}_q(\sigma)$, as can be seen from Fig.\ref{fig_cmb_wavelet_spectrum}:
\begin{equation}
\ln X^\mathrm{CMB}_q(\sigma) \simeq a_\mathrm{CMB} + s_\mathrm{CMB}\ln\sigma \,.
\label{eqn_cmb_wavelet_spec}
\end{equation}

This is due to the discrete sampling of the wavelet moments as well as the inherent statistics of the wavelet spectra of
order $q$ which can be interpreted as suitably weighted $q$-point correlation functions in Fourier analysis (compare
Sect.~\ref{analogy}).

\begin{figure}
\resizebox{\hsize}{!}{\includegraphics{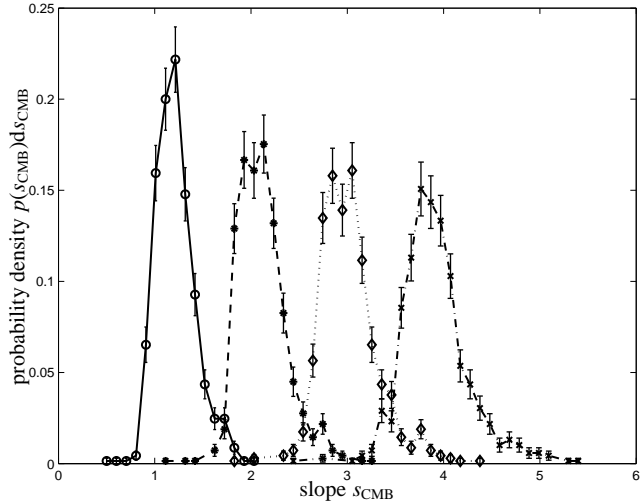}}
\caption{Distributions of the power-law slopes $s_\mathrm{CMB}$ of CMB wavelet spectra as a function of wavelet moment
order: $q=3$ (circles, solid), $q=4$ (stars, dashed), $q=5$ (diamonds, dotted), $q=6$ (crosses, dash-dotted). For the
analysing wavelet, the symlet {\em sym2} was chosen.}
\label{fig_cmb_wavelet_slope}
\end{figure}

Fig.~\ref{fig_cmb_wavelet_slope} shows the probability distribution function $p(s_\mathrm{CMB})\dd s_\mathrm{CMB}$ of
the slopes $s_\mathrm{CMB}$ following from linear fits to the wavelet moments $X_q(\sigma)$ for the range of $q^\prime$s
considered here. Again, the {\em sym2}-wavelet was chosen as the analysing wavelet. The slopes $s_\mathrm{CMB}$ are not
well confined, keeping the vast range of angular scales in mind, which in turn will make it difficult to subtract the
CMB-contribution to the wavelet spectrum of the combined map.

In order to disentangle the contributions from the CMB noise from those of the cluster, one may pursue different
approaches: Among others, CMB fluctuations underneath the cluster can be reconstructed with spline polynomials and
successively subtracted. Here, we have masked the cluster and fitted $5^\mathrm{th}$-order polynomials to the remaining
data points. Because the $y$-maps and the realisations of the CMB are to leading order combined linearly and because
the CMB is a smoothly varying field, it is possible to reconstruct the CMB fluctuations from the environment of the
cluster and interpolate to the cluster centre. The reconstructed CMB field can be subtracted from the inital image and
by applying wavelet decomposition to the cleaned field one obtains a wavelet spectrum, from which the parameters $a$,
$c$ and $s$ can be reliably derived.

\subsection{Redshift estimation}\label{ana_accuracy}
\begin{figure}
\begin{tabular}{c}
\resizebox{\hsize}{!}{\includegraphics{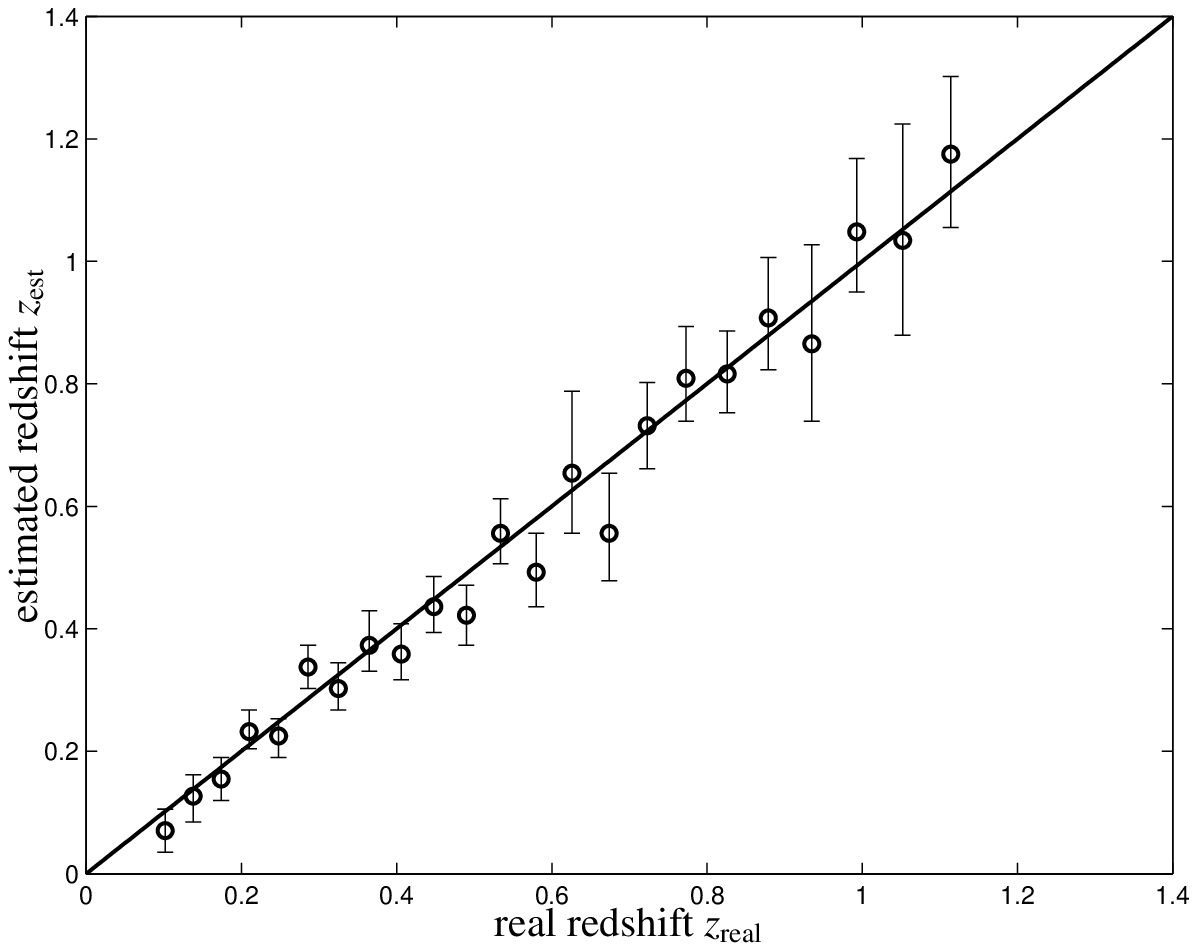}}\\
\resizebox{\hsize}{!}{\includegraphics{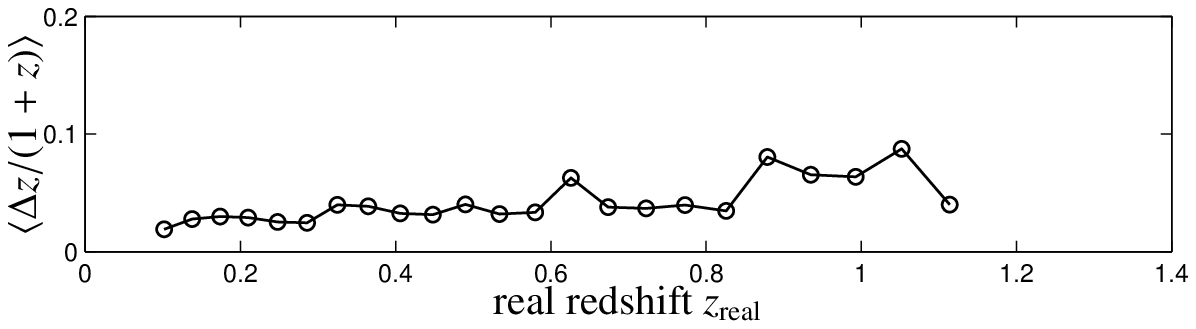}}
\end{tabular}
\caption{Redshift determination and error estimation from all three parameters $a$, $c$ and $s$ that followed from
wavelet analysis with the {\em sym2}-wavelet. The upper panel shows the estimated redshift $z_\mathrm{est}$ and its
error $\Delta z$ and the lower panel shows the relative accuracy $\Delta z / (1+z)$, both as a function of redshift
$z_\mathrm{real}$. Here, CMB fluctuations were not taken into account. The value of the wavelet moments was set to be
$q=3$.}
\label{fig_zest_err_cmb_ac}
\end{figure}

\begin{figure}
\begin{tabular}{c}
\resizebox{\hsize}{!}{\includegraphics{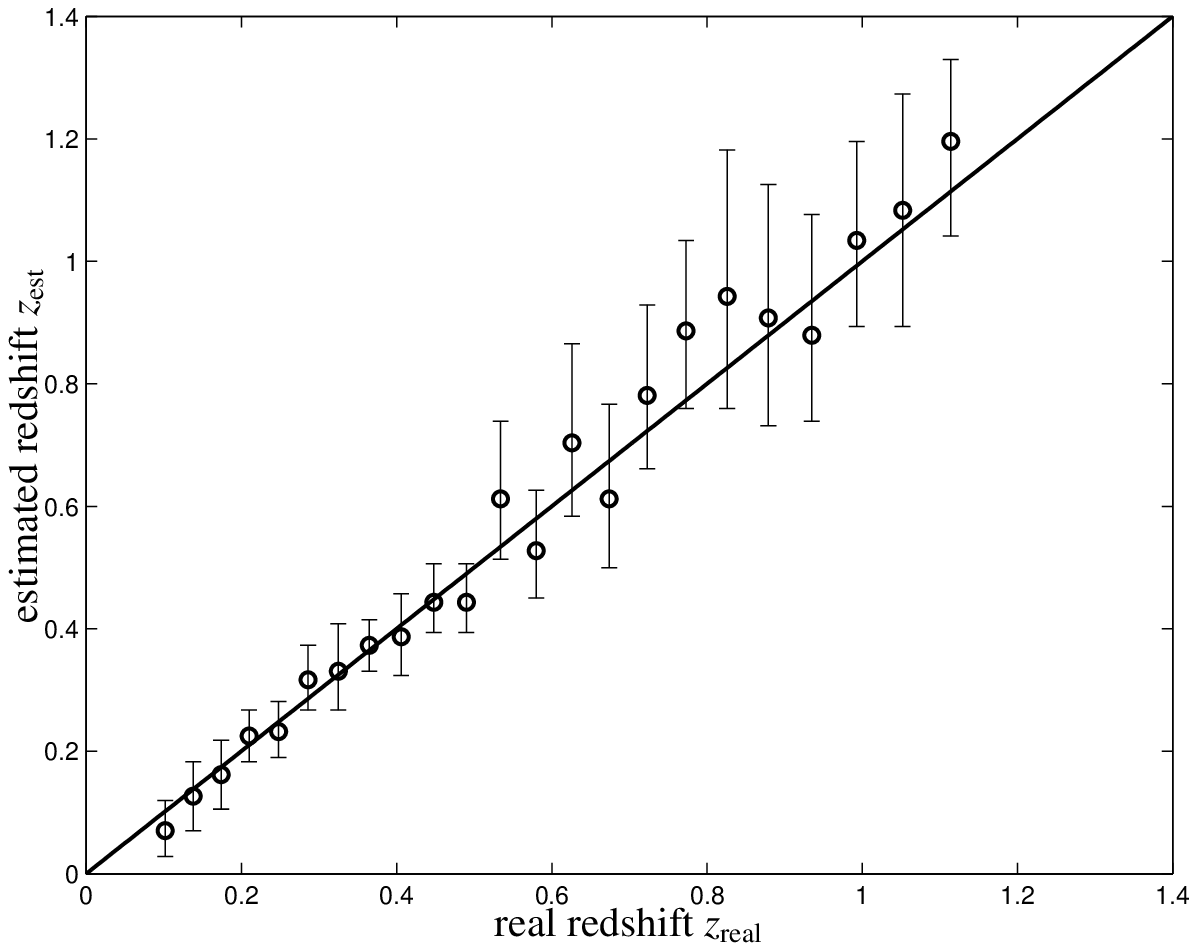}}\\
\resizebox{\hsize}{!}{\includegraphics{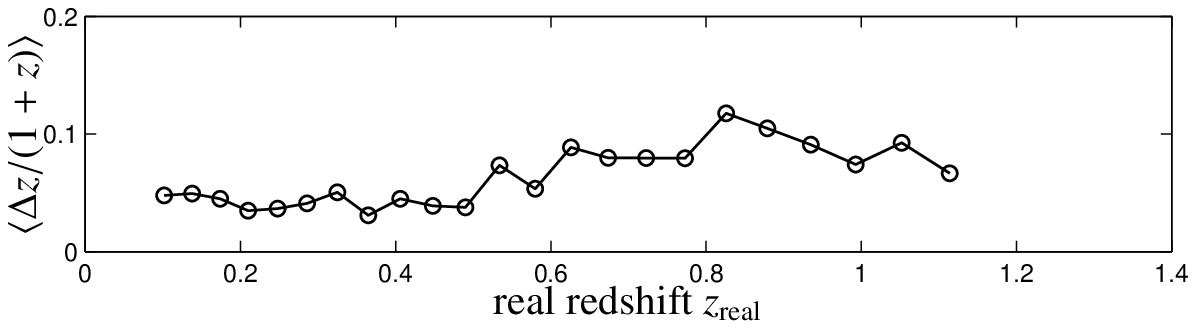}}
\end{tabular}
\caption{Redshift determination and error estimation from all three parameters $a$, $c$ and $s$ resulting from
wavelet decomposition of the combined maps (i.e. with CMB) using the {\em sym2}-wavelet. In the upper panel, the
estimated redshift $z_\mathrm{est}$ and its error $\Delta z$  is shown a function of real redshift $z_\mathrm{real}$.
In comparison, the relative accuracy $\Delta z / (1+z)$ as a function of $z_\mathrm{real}$ is shown in the lower
panel. Again, the order of the wavelet moments was taken to be $q=3$.}
\label{fig_zest_err_cmb_acs}
\end{figure}

In order to assess the accuracy of the redshift measurement, a maximum likelihood estimation is performed. The
likelihood function is defined as:
\begin{equation}
\mathcal{L}(z) =
\frac{1}{(2\,\pi)^{3/2}\, \sigma_a\,
\sigma_c\,\sigma_s}\exp\left(-\sum\limits_{x\in\left\{a,c,s\right\}}\frac{1}{N}\sum\limits_{i=1}^N\frac{(x_i-x(z))^2}{2
\sigma_x^2}\right)\mbox{,}\label{eqn_maxlikeli}
\end{equation}
and was evaluated for each bin separately, i.e. the index $i$ enumerates clusters within the redshift bin under
consideration. $N=30$ denotes the number of clusters within a single redshift bin. From the position of the maximum in
$\mathcal{L}(z)$ the most probable redshift estimate $z$ was derived and the accuracy of the estimate followed from the
percentiles corresponding to $1\mbox{-}\sigma$ confidence intervals. Fig.~\ref{fig_zest_err_cmb_ac} shows the estimated
redshift versus the real redshift for the cluster sample derived by using all of the three parameters $a$, $c$ and $s$.
In comparision, the error bars have become larger by a factor of $\simeq 1.5$ when including the fluctuating CMB, as
illustrated by Fig.~\ref{fig_zest_err_cmb_acs}. The measurement is unbiased and the error relative to $1+z$ rises
slightly with increasing redshift $z$.

The results for different analysing wavelets as a function of wavelet moment order $q$ are summarised
in tables~\ref{table_redshift_pure} and \ref{table_redshift_CMB}. Clearly, the method starts to fail at
redshifts exceeding $\gsim 1$, when the angular diameter distance $\dang(z)$ develops a plateau and does not
cause clusters to appear smaller. The average attainable accuracy is stated relative to $1+z$ in order to
facilitate comparison to photometric redshifts. The accuracy slightly degrades with increasing $q$, which is due to
suppression of small wavelet expansion coefficients especially at small scales and the resulting inaccuracy of
the fitting formula eqn.~(\ref{fitting_formula}) used to extract the spectral parameters $a$, $c$ and $s$ from
the wavelet spectrum.

\begin{table}
\vspace{-0.1 cm}
\begin{center}
\begin{tabular}{llcccc}
\hline \hline\vphantom{\Large A}%
wavelet family  	& wavelet 	& $q=3$ 	& $q=4$ 	& $q=5$ 	& $q=6$ 	\\\hline
\vphantom{\Large A}%
symlet		 	& {\em sym2}	& 4.1\%		& 4.4\%		& 4.7\%		& 4.8\%		\\
symlet			& {\em sym3}	& 4.3\%		& 4.8\%		& 5.1\%		& 5.2\%		\\
Daubechies'	    	& {\em db4}	& 5.2\%		& 5.3\%	    	& 5.4\%	   	& 5.4\%	    	\\
Daubechies'	    	& {\em db5}	& 5.5\%		& 5.0\%	    	& 4.9\%	   	& 4.8\%	    	\\
coiflet	    		& {\em coif1}	& 4.2\%		& 4.4\%	    	& 4.8\%	    	& 5.0\%	    	\\
biorthogonal    	& {\em bior1.3}	& 5.5\%		& 5.4\%	    	& 5.4\%	    	& 5.4\%	    	\\
\hline
\end{tabular}
\end{center}
\caption{Averaged accuracy of the redshift-determination relative to $1+z$ based on three parameters derived from the
wavelet spectrum of order $q$ without the noise contribution from the fluctuating CMB.}\label{table_redshift_pure}
\end{table}

Inclusion of the CMB in order to test the applicability of determining morphological redshifts in the case of
single-frequency interferometers results in a deterioration of the redshift estimation accuracy of a factor close to
1.5, which is caused by imperfections of the CMB removal by $5^{th}$-order spline polynomials.

\begin{table}
\vspace{-0.1 cm}
\begin{center}
\begin{tabular}{llcccc}
\hline \hline\vphantom{\Large A}%
wavelet family  	& wavelet 	& $q=3$ 	& $q=4$ 	& $q=5$ 	& $q=6$ 	\\
\hline
\vphantom{\Large A}%
symlet			& {\em sym2}	& 6.2\%		& 6.3\%		& 6.2\%		& 6.3\%		\\
symlet          	& {\em sym3}	& 6.7\%		& 6.5\%	  	& 6.4\%	    	& 7.2\%	    	\\
Daubechies'	    	& {\em db4}	& 6.9\%	    	& 6.8\%	  	& 6.9\%	    	& 6.9\%	    	\\
Daubechies'	    	& {\em db5}	& 7.6\%	    	& 7.4\%	  	& 7.3\%	    	& 7.2\%	    	\\
coiflet	    		& {\em coif1}  	& 6.1\%	    	& 5.9\%	 	& 6.0\%	    	& 6.8\%	    	\\
biorthogonal    	& {\em bior1.3}	& 7.5\%	    	& 7.4\%	 	& 7.2\%	    	& 7.3\%	    	\\
\hline
\end{tabular}
\end{center}
\caption{Averaged accuracy of the redshift-determination relative to $1+z$ based on three parameters derived from the
wavelet spectrum of order $q$ with  the noise contribution CMB caused by fluctuations in the
CMB.}
\label{table_redshift_CMB}
\end{table}

\section{Systematics}
\label{systematics}
SZ clusters would be self-similar and would perfectly follow scaling relations provided several requirements are
fulfilled: ({\em i}) virial equilibrium ($T\propto M^{2/3}$), ({\em ii}) structural identity, expressed in equal form
factors, ({\em iii}) a universal baryon fraction and ({\em iv}) the absence of heating and cooling processes. Each of
these assumptions may be challenged and leads to deviations from the self-similar scaling relations. While the first two
points are included in the numerical simulation and are limited by the selection criteria, they increase the scatter
in the relations between virial quantities, or equally, the wavelet parameters $a$, $c$ and $s$. Systematic trends of
the baryon fraction with cluster mass (see Sect.~\ref{sec_tilted}) and the formation of cooling flows
(Sect.~\ref{sec_cooling}) need to be assessed separately.

\subsection{Influence of tilted scaling relations}
\label{sec_tilted}
Analyses of X-ray observations carried out by \citet{1999MNRAS.305..631A} and \citet{1999ApJ...517..627M} suggest a
weak trend of the clusters baryon fraction with cluster mass $M$ and a deviation from the universal value $f_B =\Omega_b
/\Omega_m$, which is due to feedback processes like galactic winds that more effectively deplete the ICM of baryons in
low temperature clusters compared to high temperature clusters. The variation of the gas fraction, however, is
highly model dependent and reflects the mass and density estimation technique: Estimating masses by means of the virial
theorem at fixed density contrast, yields a scaling of $T\propto M^{2/3}$ and requires the baryon fraction to increase
with temperature. Alternatively, using the fit of a $\beta$-profile to the cluster, one finds the baryon fraction
$f_B$ to be nearly constant while a steeping of the slope of the $M$-$T$-relation is found.

In either of the above mentioned cases the dependence especially of the wavelet parameter $a$, which is a logarithmic
measure of the SZ flux $\Upsilon$ would be increased in more massive clusters and would thus increase the scatter in
$a$ of a cluster sample at a given redshift. The quoted analyses of X-ray data find the baryon fraction to show a
relative variation amounting to $\simeq$ 10\% at fixed temperature, i.e. at fixed depth of the potential well for a
sample of local clusters. Apartfrom the the systematic component, that can in principle beremoved, once high quality
X-ray data will improve our understanding of this phenomenon and allows proper modelling, the stochastic contribution
can only be constrained to be at most of equal relative influence to $\Delta\Upsilon/\Upsilon$ as the scatter in
morphology.

\subsection{Cooling flow clusters}
\label{sec_cooling}
In order to estimate the accuracy of the method outlined above, so far we only used adiabatic hydrodynamical simulations
which lack of cooling processes. Thus we need to address the influence of cooling flow clusters on our
proposed method. After an analytical investigation following Sect.~\ref{application} we compare clusters with and
without cooling flows and show how the morphological changes in cooling flow clusters impacts on the
wavelet spectra.

\subsubsection{Analytical wavelet transform of cooling flow clusters}
Instead of a single King profile we assume that the SZ emission of a cooling flow cluster can be described by a double
King profile for reasons of analytical feasibility:

\begin{equation}
y(\bmath{x}) = y(r) = \sum_{i=1}^2 y_i\left[1+\left(\frac{r}{r_i}\right)^2\right]^{-1},
\end{equation}
where the second term describes the additional enhancement owing to the cooling flow.  Deprojecting this two-dimensional
profile by means of \citet{2003A&A...P} yields:

\begin{equation}
p_\e(R) = \frac{1}{2}\, n_\e(R)\, k T_\e(R) =
\frac{m_\e c^2}{\sigma_\mathrm{T}} \sum_{i=1}^2 \frac{y_i}{2\pi r_i}
\frac{\mathcal{B}\left(\frac{1}{2},\frac{3}{2}\right)}{\left(1+R^2/r_i^2\right)^{3/2}},
\end{equation}
where $R$ denotes the three-dimensional radius and $\mathcal{B}(a,b)$ denotes the $\beta$-function
\citep{1965hmfw.book.....A}. Thus we obtain for the ratio of the central values of the Comptonisation parameters $y_i$

\begin{equation}
\frac{y_2}{y_1} = \frac{p_2\,r_2}{p_1\, r_1} \sim \frac{1}{2},
\end{equation}
where we inserted typical values for cooling flow clusters, $p_2/p_1 \sim 3$ and $r_2 /r_1 \sim 1/6$. The second order
wavelet moment of cooling flow clusters can be obtained in analogy to the non-cooling flow case:

\begin{equation}
  X_2^\mathrm{CF}(\sigma) = 2\pi \int \dd k\, k^5 \exp(-k^2 \sigma^2)
  \left| y_1 r_1^2 K_0(k r_1) + y_2 r_2^2 K_0(k r_2) \right|^2.
\end{equation}

This second order wavelet moment shows an increasing amplitude and a decreasing cutoff parameter compared to the one
without a cooling flow.

\subsubsection{Numerical analysis}
\begin{figure}
\resizebox{\hsize}{!}{\includegraphics{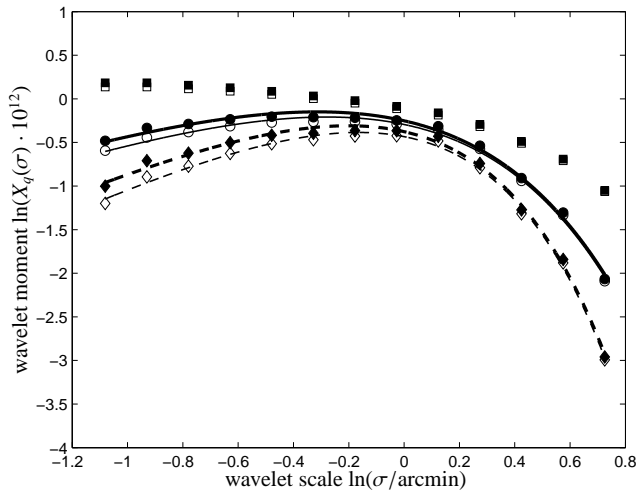}}
\caption{The influence of cooling flows to the spectrum of wavelet moments, together with the fitting
formula~(\ref{fitting_formula}) for increasing  wavelet moment order $q$: $q = 2$ (squares), $q =3$ (circles, solid),
and $q  = 4$ (diamonds, dashed) for a single cluster without instrumental smoothing. Open symbols are values
derived from the simulated non-cooling flow cluster, whereas filled symbols denote the corresponding cooling flow
cluster.}
\label{fig_cooling_flow_cluster}
\end{figure}

In order to scrutinise these findings we apply our method to adiabatically simulated clusters to which we add an
enhanced emission to mimic the SZ emission of the cooling flow. In Fig.~\ref{fig_cooling_flow_cluster} the
resulting spectra of wavelet moments are shown together with the fitting formula eqn.~(\ref{fitting_formula}) for
increasing wavelet moment order $q$.

It can clearly be seen in Fig.~\ref{fig_cooling_flow_cluster} that the enhanced emission due to the cooling flow yields
a slightly higher amplitude of the wavelet spectrum on small scales. Extracting information from the wavelet spectrum
by means of eqn.~(\ref{fitting_formula}) reveals slightly higher values for the amplitude $a$ and smaller values for
the cutoff $c$ on the percent level. However, this influence is minimised when considering finite instrumental
resolution particularly for high redshift clusters. In any case, if a prominent cooling flow is sufficiently well
resolved it could be masked and replaced by an interpolation in between the mask boundaries.

\section{Redshift estimation in a nutshell}\label{nutshell}
This section shall provide a short summary of how to apply our method to an SZ survey for estimating redshifts provided
a temperature map of a patch on the sky with resolved images of clusters.

\begin{itemize}
\item{Once a cluster candidate has been localised at a particular position of the map this cluster and its ambient field
has to be cut out. If the number of grid points amounts below $64^2$ sampling points, the mesh should be refined by
interpolation in order to reach dynamical range of approximately two decades. This is important in order to provide a
sufficiently broad range of scales to be probed by the wavelet decomposition.}

\item{The wavelet spectrum of the map is obtained by wavelet transforming the map preferentially using the {\em symlet}
basis functions (compare Sect.~\ref{definitions}). The morphological information contained within the wavelet spectrum
can be extracted by means of the model function of eqn.~(\ref{fitting_formula}). In the case of single-frequency
observations the ambient CMB field cannot be separated from the SZ signal of a cluster. The method described in
Sect.~\ref{ana_cmb_infl} might be applied in order to reconstruct the wavelet spectrum of the pure SZ cluster signal.}

\item{The redshift dependence of the wavelet parameters (amplitude $a$, cutoff $c$, and slope $s$) follows the
functional form of eqn.~(\ref{wavelet_par}). However, the single model parameters depend on the definitions of the
particular wavelets and the details of the survey, including different sources of noise and the cluster detection
criteria. The most promising way of determining the parameters of the gauge functions laid down in
eqn.~(\ref{wavelet_par}) would be to derive them from a learning set of clusters with known (photometric) redshifts. The
final redshift estimate of the cluster is most conveniently determined by means of maximum likelihood analysis, as
described by eqn.~(\ref{eqn_maxlikeli})}.
\end{itemize}

\section{Summary}\label{sum}
In this paper, a method of estimating the redshift of a cluster based on the wavelet decomposition of its resolved SZ
morphology is presented. From a fit to the spectrum of wavelet moments three spectral parameters are derived, that in
turn are non-degenerate and indicative of cluster distance. These parameters are utilized, through a maximum likelihood
technique, for estimating the cluster's redshift. In the maximum likelihood technique, empirical gauge functions
describing the wavelet parameter's $z$-dependence are used.

First, the method was tested on a simple analytical case:
The spectrum of Mexican-hat wavelet moments can be derived analytically for a King-profile, which is known to describe
the Compton-$y$ amplitude of clusters well. The spectrum of wavelet coefficients as a function of wavelet scale
$\sigma$, exhibits a break at the cluster scale $r_c$ and may thus serve as a measure of the cluster's
size. Additionally, the asymptotic behaviour of the wavelet spectrum in the limit of $\sigma\gg r_c$ and $\sigma\ll r_c$
can be understood. The derivation of wavelet moments of order $q=2$ is analogous to considering the Fourier power
spectrum of the Compton-$y$ map, filtered with Fourier transformed wavelet. The shape of the spectrum of wavelet moments
of order $q=2$ from the analytic calculation is consistent with one obtained by applying wavelet decomposition to
simulated  SZ cluster maps.

The method was then applied to set of numerically simulated SZ clusters with $1\arcmin$ (FWHM) resolution -- comparable
to the resolution of future SZ experiments. The sample comprises 690 cluster maps distributed in 23 redshift bins, which
is a comparably large cluster sample. The clusters are chosen such that they are not in a merging state and their SZ
image is not too elongated, two criteria that favour clusters close to virialisation. Additionally, in order to
simulate single-frequency observations, the cluster maps were combined with realisations of the CMB that constitute the
main source of noise.

The method was tested for a range of wavelet functions (e.g., {\em symlet, coiflet, Daubechies, biorthogonal}).
The average attainable accuracy in estimating redshifts is found to be almost independent of the specific functional
form used, although the {\em symlet} basis yielded the best results. However, the method could benefit from
improvements concerning the choice of the wavelet basis. For instance, one could try to construct an optimised wavelet
specifically for $\beta$-profiles, that yields maximised wavelet coefficients $\chi(\bmath{\mu},\sigma)$.

As expected, there is only a weak change in accuracy with respect to the order $q$ of the chosen wavelet
moment $X_q(\sigma)$. This, however, is most likely to change when applying the wavelet analysis to noisy images,
because for increasing choices of $q$, uncorrelated noise is suppressed relative to the cluster's signal
and concentrating on higher values for $q$ should provide a more robust measurement of the set of structural parameters
$a$, $c$ and $s$. The increment of $q$ itself is limited by numerics -- this is the case when the
wavelet moment $X_q(\sigma)$ is dominated by the largest wavelet expansion coefficient $\chi(\bmath{\mu},\sigma)$, and
does not reflect anymore the dependence on the wavelet scale $\sigma$. In this limit, the wavelet spectrum would exhibit
a generic power law behaviour: $X_q(\sigma)\propto\sigma^{\gamma(q)}$ for large $q$. The structural parameters $a$, $c$
and $s$ were found to depend on redshift $z$ by a simple exponential (eqn.~(\ref{wavelet_par})). The free parameters in
this equation can be determined from a (relatively small) sample of SZ cluster images with known redshift.

The accuracy of determining cluster distances has been assessed by maximum likelihood estimation. The method yields
accuracies of $4 - 5\%$ relative to $1+z$, which is competitive with photometric redshifts, but reaches out to
larger distances. At redshifts exceeding $z\gsim 1$, the accuracy is expected to degrade because the angular
diameter distance $\dang(z)$ starts to level off and thus sets the limit of applicability. For single frequency data,
the CMB fluctuations can be removed with a simple polynomial reconstruction approach; the accuracy in the redshift
estimation is then decreased to $6 - 7\%$.

In this work we have considered two major systematic effects that might degrade the accuracy of the method. The first is
the varying baryon fraction with cluster mass, which has been studied only for local cluster samples. While the
systematic trend could in principle be corrected for, the stochastic contribution will always add to the uncertainty of
the distance determination. Another systematic is the influence of cooling flows at the cluster's centre. In this case
we have been able to show that the uncertainty it adds to the redshift estimate is very small, mainly because the volume
occupied by the cooling flow region is limited to the cluster's core.

Although the result in the distance estimation is stated in terms of redshift, it should be emphasised that a
specific cosmology is assumed, which is needed for converting the observables, namely the wavelet parameters, to a
distance estimate. The distances following from the analysis have been expressed as redshifts because of their
elementary interpretation, but the implicit assumption of an underlying cosmology should be kept in mind when comparing
to e.g. photometric redshifts. For that reason, the precision of the method presented is limited by the accuracy to
which the cosmological parameters are known. Apart from being a distance indicator, the redshift also plays the role
of an evolutionary parameter.

Comparing this work to the pioneering paper by \citet{2003MNRAS.341..599D}, our expectations concerning the accuracy
of morphological redshifts are even more optimistic: Without fitting $\beta$-profiles to the observational data, it is
possible to describe the cluster's SZ morphology by solely relying on wavelet decomposition.  Also, we describe the
spectrum of wavelet moments with a small set of structural parameters, that have a lucid physical interpretation,
provide a non-degenerate distance measurement and enable redshift determination owing to their monotonic decline with
redshift. Furthermore, the redshift dependence of the structural parameters is calibrated with the data set itself
without relying on {\em prior} and simplifying assumptions. In spite of the small number of observables considered here,
the accuracy in the redshift estimation of this method is doubled, in comparison with \citet{2003MNRAS.341..599D}, even
for a single frequency experiment.

\section*{Acknowledgements}
The authors would like to thank Volker Springel and Lars Hernquist who kindly provided us with their numerical
simulations and Antonaldo Diaferio, Matthias Bartelmann and Simon D.M. White for clarifying discussions and many useful
comments.

\bibliography{bibtex/references}
\bibliographystyle{bibtex/aa}

\appendix

\bsp

\label{lastpage}

\end{document}